\documentclass[
,twocolumn%
 ,secnumarabic%
,amssymb, amsmath,nobibnotes, aps, prb,superscriptaddress]{revtex4}

\usepackage{graphics}
\usepackage{bm}

\begin{document}

\title{Microwave photovoltage and
photoresistance effects in ferromagnetic microstrips}
\author{N. Mecking}
\email{nmecking@physnet.uni-hamburg.de} \affiliation{Department of
Physics and Astronomy, University of Manitoba, Winnipeg, Canada R3T
2N2} \affiliation{Institut f\"ur angewandte Physik und Zentrum f\"ur
Mikrostrukturforschung, Universit\"at Hamburg, Jungiusstra\ss e 11,
20355 Hamburg, Germany}
\author{Y.S. Gui}
\author{C.-M. Hu}
\email{hu@physics.umanitoba.ca} \affiliation{Department of Physics
and Astronomy, University of Manitoba, Winnipeg, Canada R3T 2N2}

\begin{abstract}
We investigate the dc electric response induced by ferromagnetic
resonance in ferromagnetic Permalloy (Ni$_{80}$Fe$_{20}$)
microstrips. The resulting magnetization precession alters the angle
of the magnetization with respect to both dc and rf current.
Consequently the time averaged anisotropic magnetoresistance (AMR)
changes (photoresistance). At the same time the time-dependent AMR
oscillation rectifies a part of the rf current and induces a dc
voltage (photovoltage). A phenomenological approach to
magnetoresistance is used to describe the distinct characteristics
of the photoresistance and photovoltage with a consistent formalism,
which is found in excellent agreement with experiments performed on
in-plane magnetized ferromagnetic microstrips. Application of the
microwave photovoltage effect for rf magnetic field sensing is
discussed.
\end{abstract}

\maketitle

\section{INTRODUCTION}
\label{intro}

The fact that macroscopic mutual actions exist between electricity
and magnetism has been known for centuries as described in many text
books of electromagnetism \cite{Guru}. Now, this subject is
transforming onto the microscopic level, as revealed in various
spin-charge coupling effects studied in the new discipline of
spintronics. Among them, striking phenomena are the dc charge
transport effects induced by spin precession in ferromagnetic
metals, which feature both academic interest and technical
significance \cite{Gurney,Zhu}. Experiments have been performed
independently by a number of groups on devices with different
configurations
\cite{Tulapurkar,Sankey,Azevedo,Saitoh,Costache,Costache06,Gui,GuiHu,Gui2,Gui07c,Yamaguchi,Oh,Goennenwein}.
Most works were motivated by the study of spin torque
\cite{Slonczewski,Berger96}, which describes the impact of a
spin-polarized charge current on the magnetic moment. In this
context, Tulapurkar \emph{et al.} made the first spin-torque diode
\cite{Tulapurkar}, and Sankey \emph{et al.} detected the
spin-torque-driven ferromagnetic resonance (FMR) electrically
\cite{Sankey}. Both measured the vertical transport across
nano-structured magnetic multilayers. Along a parallel path, a
number of works \cite{Berger99,Brataas,Wang} have been devoted to
study the effect of spin pumping. One of the interesting predictions
is that injection of a spin current from a moving magnetization into
a normal metal induces a dc voltage across the interface. To detect
such a dc effect induced by spin pumping\cite{Brataas}, experiments
have been performed by measuring lateral transport in hybrid devices
under rf excitation \cite{Azevedo,Saitoh,Costache}.

From a quite different perspective, Gui \emph{et al.} set out to
explore the general impacts of the high frequency response on the dc
transport in ferromagnetic metals \cite{Gui}, based on the
consideration that similar links in semiconductors have been
extensively applied for electrical detection of both spin and charge
excitations \cite{Hu}. Gui \emph{et al.} detected, subsequently,
photoresistance induced by bolometric effect \cite{Gui}, as well as
photocurrent \cite{GuiHu}, photovoltage \cite{Gui2}, and
photoresistance \cite{Gui07c} caused by the spin-rectification
effect. A spin dynamo \cite{GuiHu} was thereby realized for
generating dc current via the spin precession, and the device was
applied for a comprehensive electrical study of the characteristics
of quantized spin excitations in micro-structured ferromagnets
\cite{Gui2}. The spin-rectification effect was independently
investigated by both, Costache \emph{et al.} \cite{Costache06} and
Yamaguchi \emph{et al.} \cite{Yamaguchi}, and seems to be also
responsible for the dc effects detected earlier by Oh \emph{et al.}
\cite{Oh}. A method for distinguishing the photoresistance induced
by either spin precession or bolometric effect was recently
established \cite{Gui07c}, which is based on the nice work performed
by Goennenwein \emph{et al.} \cite{Goennenwein}, who determined the
response time of the bolometric effect in ferromagnetic metals.

While most of these studies, understandably, tend to emphasize
different nature of dc effects investigated in different devices, it
is perhaps more intriguing to ask the questions whether the
seemingly diverse but obviously related phenomena could be described
by a unified phenomenological formalism, and whether they might
arise from a similar microscopic origin. From a historical
perspective, these two questions reflect exactly the spirit of two
classic papers \cite{Juretschke,Silsbee} published by Juretscheke
and Silsbee \emph{et al.}, respectively, which have been often
ignored but have shed light on the dc effects of spin dynamics in
ferromagnets. In the approach developed by Juretscheke, photovoltage
induced by FMR in ferromagnetic films was described based on a
phenomenological depiction of magnetoresistive effects
\cite{Juretschke}. While in the microscopic model developed by
Silsbee \emph{et al.} based on the combination of Bloch and
diffusion equations, a coherent picture was established for the spin
transport across the interface between ferromagnets and normal
conductors under rf excitation \cite{Silsbee}.

The goal of this paper is to provide a consistent view for
describing photocurrent, photovoltage, and photoresistance of
ferromagnets based on a phenomenological approach to
magnetoresistance. We compare the theoretical results with
experiments performed on ferromagnetic microstrips in detail. The
paper is organized in the following way: in section 2, a theoretical
description of the photocurrent, photovoltage, and photoresistance
in thin ferromagnetic films under FMR excitation is presented.
Sections 2.1-2.4 establish the formalism for the microwave
photovoltage (PV) and photoresistance (PR) based on the
phenomenological approach to magnetoresistance. These arise from the
non-linear coupling of microwave spin excitations (resulting in
magnetization $\mathbf M$ precession) with charge currents by means
of the anisotropic magnetoresistance (AMR). Section 2.5 compares our
model with the phenomenological approach developed by Juretscheke.
Section 2.6 provides a discussion concerning the microwave
photovoltage and photoresistance based on other magnetoresistance
effects (like anomalous Hall effect (AHE), giant magnetoresistance
(GMR) and tunnelling magnetoresistance (TMR)).

Experimental results on microwave photovoltage and photoresistance
measured in ferromagnetic microstrips are presented in section
\ref{voltmeas} and \ref{resmeas}, respectively. We focus in
particular on their characteristic different line shapes, which can
be well explained by our model. In section \ref{conclu} conclusions
and an outlook
are given.\\

\section{MICROWAVE
PHOTOVOLTAGE AND PHOTORESISTANCE BASED ON PHENOMENOLOGICAL AMR}
\label{theory}

\subsection{AMR COUPLING OF SPIN AND CHARGE}
\label{amr}

The AMR-coupling of spin and charge in ferromagnetic films results
in microwave photovoltage and photoresistance. The photovoltage can
be understood regarding Ohms law (current $I(t)$ and voltage $U(t)$)

\begin{equation}
U(t)=R(t)\cdot I(t) \label{eq:OhmsLaw}
\end{equation}

\begin{figure}
\includegraphics{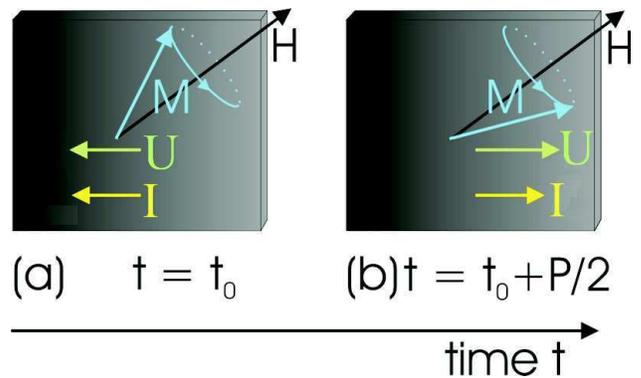}
\caption{(Color online). Mechanism of the AMR-induced microwave
photovoltage: $\mathbf M$ precesses (period P) in phase with the rf
current $\mathbf I$. (a) $\mathbf M$ lying almost perpendicular to
$\mathbf I$ results in low AMR. (b) $\mathbf M$ lying almost
parallel to $\mathbf I$ results in high AMR. The time average
voltage $\mathbf U$ becomes non-zero.} \label{fig:dceffect}
\end{figure}

We consider a time-dependent resistance $R(t)=R^0+R^1\cos(\omega
t-\psi)$ which oscillates at the microwave frequency $\omega=2\pi f$
due to the AMR oscillation arising from magnetization precession.
$\psi$ is the oscillations phase shift with respect to the phase of
the rf current $I(t)$. For the sake of generality $\psi$ will be
kept as a parameter in this work and will be discussed in detail in
section \ref{asymm}. $I(t)$ takes the form $I(t)=I_1\cos(\omega t)$
and is induced by the microwaves. It follows that $U(t)$ consists of
time-dependent terms with the frequency $\omega$, $2\omega$ and a
constant term (time independent) which corresponds to the time
average voltage and is equal to the photovoltage:
$U_{MW}=\left<R^1I_1\cos(\omega t-\psi)\cos(\omega t)\right>
=(R^1I_1\cos\psi)/2$ ($\left<\ \right>$ denotes time-averaging). A
demonstrative picture of the microwave photovoltage mechanism can be
seen in figure \ref{fig:dceffect}.

\begin{figure}
\centering
\includegraphics{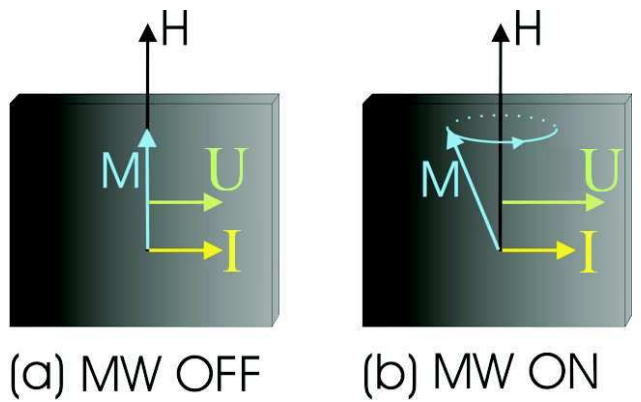}
\caption{(Color online). Mechanism of the AMR-induced
photoresistance: (a) Without microwaves (MW) $\mathbf M$ lies
perpendicular to the dc current $\mathbf I$ and the AMR is minimal
(b) With microwaves $\mathbf M$ precesses and is not perpendicular
to $\mathbf I$ anymore. Consequently the AMR increases (higher
voltage drop $\mathbf U$).} \label{fig:RAlteration}
\end{figure}

The second effect we investigate which is also based on AMR
spin-charge coupling is the microwave photoresistance $\Delta
R_{MW}$. This has been reported recently \cite{Costache06} with the
equilibrium magnetization $\mathbf M_0$ of a ferromagnetic stripe
aligned to a dc current $\mathbf I_{0}$. Microwave induced
precession then misalignes the dynamic magnetization $\mathbf M$
with respect to $\mathbf{I_{0}}$ and thus makes the AMR drop
measurably. In this work we present results which also show that if
$\mathbf M_0$ lies perpendicular to $\mathbf I_{0}$ the opposite
effect takes place: Microwave induced precession causes $\mathbf M$
to leave its perpendicular position what increases the AMR (see
figure \ref{fig:RAlteration}).

\begin{figure}
\centering
\includegraphics{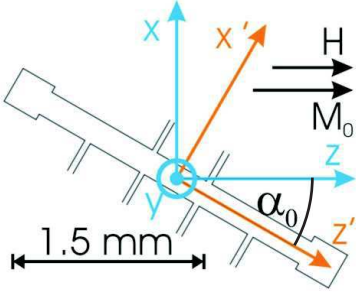}
\caption{(Color online). (x,y,z) and (x$^\prime$,y,z$^\prime$)
coordinate systems in front of a layout of our Permalloy film stripe
($200\times2400\ \mu$m$^2$) with 2 contacts and 6 side junctions.}
\label{fig:koordinaten}
\end{figure}

After this qualitative introduction we want to go ahead with a
quantitative description of the AMR induced microwave photovoltage
and photoresistance. Therefore we define an orthogonal coordinate
system (x,y,z) (see figure \ref{fig:koordinaten}). The y-axis lies
normal to the film plane and the z-axis is aligned with the magnetic
field $\mathbf H$ and hence with the magnetization $\mathbf M$ which
is always aligned with $\mathbf H$ in our measurements because of
the sample being always magnetized to saturation.

Geometrically our samples are thin films patterned to stripe shape,
so that $d\ll w\ll l$, where $d$, $w$ and $l$ are the thickness,
width and length of the sample. We apply $\mathbf H$ always in the
ferromagnetic film plane. For calculations based on the stripes
geometry the coordinates $x^\prime$ and $z^\prime$ are defined.
These lie in the film plane. $x^\prime$ is perpendicular and
$z^\prime$ parallel to the stripe. The following coordinate
transformation applies:
$(x,y,z)=(x^\prime\cos(\alpha_0)-z^\prime\sin(\alpha_0),
y,z^\prime\cos(\alpha_0)+x^\prime\sin(\alpha_0))$ where $\alpha_0$
is the angle between $\mathbf H$ and the stripe.

\begin{figure}
\centering
\includegraphics{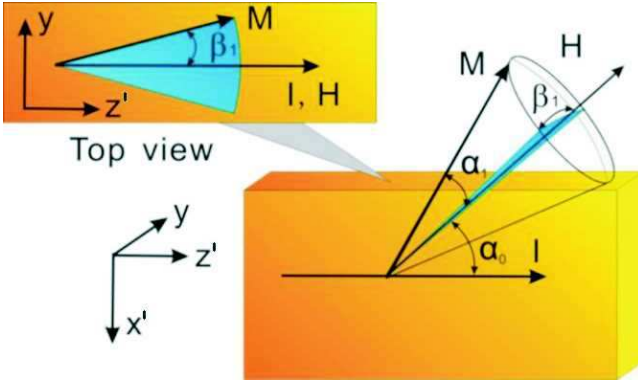}
\caption{(Color online). Sketch of the magnetization precession. The
magnetic field $\mathbf H$ encloses the angle $\alpha_0$ with the
current $\mathbf I$. The magnetization oscillation towards $\mathbf
I$ has the amplitude $\alpha_1$ and that perpendicular to $\mathbf
I$: $\beta_1$. } \label{fig:Msphere}
\end{figure}

For the microwave photovoltage and photoresistance the longitudinal
resistance $R(t)=R_0+R_A\cos^2\theta(t)$ of the film stripe matters.
It consists of the minimal longitudinal resistance $R_0$ and the
additional resistance $R_A\cos^2\theta(t)$ from AMR. $\theta(t)$ is
the angle between the $z^\prime$-axis (parallel to the stripe) and
$\mathbf M$. $\mathbf M$ moves on a sphere with the radius $M_0$,
which is the saturation magnetization of our sample. $\theta(t)$ can
be decomposed into the angle $\alpha(t)$ in the ferromagnetic film
plane and the out-of-plane angle $\beta(t)$ (see figure
\ref{fig:Msphere}). Consequently:

\begin{equation}
\cos{\theta(t)}=\cos{\alpha(t)}\cos{\beta(t)} \label{eq:costheta}
\end{equation}

Precession of the magnetization then yields oscillation of
$\alpha(t)$, $\beta(t)$ and $\theta(t)$. In our geometry the
equilibrium magnetization $\mathbf{M_0}$ encloses the in-plane
angle $\alpha_0$ with the stripe. Hence in time average
$<\beta(t)>\ =0$ and $<\alpha(t)>\ =\alpha_0$. In general the
magnetization precession is elliptical. Its principle axis lie
along the x- and y-axis and correspond to the amplitudes
$\alpha_1$ and $\beta_1$ of the in- and out-of-plane angles
$\alpha_1^t$ and $\beta_1^t$ of the rf magnetization:
$\alpha(t)=\alpha_0+\alpha_1^t(t)= \alpha_0+\alpha_1\cos(\omega
t-\psi)$ and $\beta(t)=\beta_1^t(t)=-\beta_1\sin(\omega t-\psi)$
(see figure \ref{fig:Msphere}). Using equation (\ref{eq:costheta})
we approximate $\cos^2\theta(t)$ to second order in $\alpha_1^t$
and $\beta_1^t$:

\begin{eqnarray}
\cos^2\theta(t) \approx \cos^2\theta|_{\alpha_1^t=\beta_1^t=0}+
\alpha_1^t\cdot\frac{d\cos^2\theta}{d\alpha_1^t}|_{\alpha_1^t=\beta_1^t=0}
+0\nonumber\\
+\frac{\alpha_1^{t2}}2\cdot\frac{d^2\cos^2\theta}{d\alpha_1^{t2}}|_{\alpha_1^t=\beta_1^t=0}
+\frac{\beta_1^{t2}}2\cdot\frac{d^2\cos^2\theta}{d\beta_1^{t2}}|_{\alpha_1^t=\beta_1^t=0}
\end{eqnarray}

The first order in $\beta_1^t$ vanishes because it is proportional
to $(\sin\beta)|_{\beta_1=0}=0$. It follows:

\begin{eqnarray}
\cos^2\theta(t)\approx\cos^2\alpha_0-\alpha_1\cdot\sin2\alpha_0\cos(\omega
t-\psi)\nonumber\\ -\alpha_1^2\cdot\cos2\alpha_0\cos^2(\omega
t-\psi) -\beta_1^2\cdot\cos^2\alpha_0\sin^2(\omega t-\psi)
\label{eq:theta}
\end{eqnarray}

This equation is now used to calculate the longitudinal stripe
voltage. To consider the general case an externally applied dc
current $I_0$ and a microwave induced rf current $I_1$, are included
in $I(t)=I_0+I_1\cos(\omega t)$. It follows from equation
(\ref{eq:OhmsLaw}):

\begin{equation}
U(t) = (R_0+R_A\cos^2\theta(t))\cdot(I_0+I_1\cos(\omega t))
\label{eq:dynOhm}
\end{equation}

Consequently $U(t)$ can be written as $U(t)=U_0+U_1\cos(\omega
t-\psi_1)+U_2\cos(2\omega t-\psi_2) +U_3\cos(3\omega t-\psi_3)$. For
the photovoltage and photoresistance only the constant term $U_0$,
which is equivalent to the time average voltage $\left<U(t)\right>$,
matters. Combining equation (\ref{eq:theta}) and (\ref{eq:dynOhm}),
we find:

\begin{eqnarray}
U_0=I_0(R_0+R_A\cos^2\alpha_0)
-I_1R_A\alpha_1\sin2\alpha_0\cos(\psi)/2\nonumber\\
-I_0(\alpha_1^2\cos2\alpha_0+\beta_1^2\cos^2\alpha_0)R_A/2
\label{eq:RMW}
\end{eqnarray}

Note that: $\left<\sin^2(\omega t-\psi)\right>=\left<\cos^2(\omega
t-\psi)\right>=1/2$, and $\left<\cos\omega t\cos(\omega
t-\psi)\right>=\cos(\psi)/2$. The first term in equation
(\ref{eq:RMW}) is independent of the rf quantities $I_1$, $\alpha_1$
and $\beta_1$ and represents the static voltage drop of $I_0$. The
second term is the microwave photovoltage $U_{MW}$. It shows no
impact from the dc current $I_0$. The third term represents the
microwave photoresistance $\Delta R_{MW}$. It is proportional to
$I_0$ and depends on the microwave quantities $\alpha_1$ and
$\beta_1$. By the way: It can be seen now that the rf resistance
amplitude $R^1$ used in the beginning of this paragraph corresponds
to: $R^1=R_A\alpha_1\sin 2\alpha_0$.

To analyze the magnetization's angle oscillation amplitudes
$\alpha_1$ and $\beta_1$ it is necessary to express them by means of
the corresponding rf magnetization $\Re(\mathbf m e^{-i\omega t})$.
$\mathbf m$ is the complex rf magnetization amplitude. Its phase is
defined with respect to $I_1$, so that $\Re(m_xe^{-i\omega t})$ is
in phase with $I_1\cos\omega t$ at the FMR. Because $\mathbf
M=\mathbf{M_0}+\mathbf m$, $\mathbf m=(m_x,m_y,0)$ can (in first
order approximation) only lie perpendicular to $\mathbf M_0$ because
$\mathbf M$ and $\mathbf{M_0}$ have the same length ($M_0$). Hence
$|m_x|/M_0 = \sin\alpha_1 \approx \alpha_1$ and $|m_y|/M_0 =
\sin\beta_1 \approx \beta_1$ for $\alpha_1,\beta_1\ll90^\circ$.

The microwave photovoltage and photoresistance appear whenever
magnetization precession is excited. This means if the microwaves
are in resonance with the FMR, with standing exchange spin waves
perpendicular to the film \cite{Moller,GuiHu,Gui2} or with
magnetostatic modes \cite{Gui2}. In this article we will analyze the
FMR induced microwave photoresistance and photovoltage.

\subsection{MAGNETIZATION DYNAMICS}
\label{magnet}

To understand the impact of the applied rf magnetic field
$\Re(\mathbf h e^{-i\omega t})$ on the microwave photovoltage and
photoresistance the effective susceptibilities $\chi_{xx}$,
$\chi_{xy}$ and $\chi_{yy}$, which link $\mathbf me^{-i\omega t}$
inside the sample with the complex external rf magnetic field
$\mathbf he^{-i\omega t}=(h_x,h_y,h_z)e^{-i\omega t}$ outside the
sample, have to be calculated. Here $\psi$ is encoded in the complex
phase of $\mathbf m$.

The susceptibility inside the sample (magnetic field
$\mathbf{h^{in}}e^{-i\omega
t}=(h_x^{in},h_y^{in},h_z^{in})e^{-i\omega t}$) is determined by the
Polder tensor\cite{Polder} $\hat\chi$ (received from solving the
Landau-Liftshitz-Gilbert equation \cite{LLG}):

\begin{equation}
\mathbf m=\hat\chi\mathbf{h^{in}} =
\left(\begin{array}{ccc} \chi_L & i\chi_T & 0\\
-i\chi_T& \chi_L & 0\\
0      &    0   &0
\end{array}\right)\mathbf{h^{in}}
\end{equation}

with

\begin{eqnarray}
\chi_L = \frac{\omega_M\omega_r}{\omega_r^2-\omega^2}\nonumber,\
\chi_T = \frac{\omega\omega_M}{\omega_r^2-\omega^2}\nonumber
\end{eqnarray}

where $\omega_M=\gamma M_0$ with the gyromagnetic ratio
$\gamma\approx \mu_0\cdot e/m=2\pi\mu_0\cdot28$ GHz/T (electron
charge $e$ and mass $m_e$) and $\omega_r=\gamma H$ without damping.
Approximation of our sample as a 2 dimensional film results in the
boundary conditions that $h_x$ and $b_y$ are continuous at the film
surface meaning $h_x=h_x^{in}$ and $b_y=\mu_0 h_y =
\mu_0((1+\chi_{L})h_y^{in}-i\chi_Th_x^{in})$. Hence:

\begin{equation}
\mathbf m = \left(\begin{array}{ccc} \chi_{xx} & i\chi_{xy} & 0\\
-i\chi_{xy}& \chi_{yy} & 0\\
0      &    0   &0
\end{array}\right)\mathbf h
\label{eq:Polder}
\end{equation}

with

\begin{eqnarray}
\chi_{xx} =
\frac{\omega_r\omega_M+{\omega_M}^2}{\omega_r\left(\omega_r+\omega_M\right)-\omega^2}
\nonumber\\
\ \chi_{xy} =
\frac{\omega\omega_M}{\omega_r\left(\omega_r+\omega_M\right)-\omega^2}\nonumber
\\
\chi_{yy} =
\frac{\omega_r\omega_M}{\omega_r\left(\omega_r+\omega_M\right)-\omega^2}\nonumber
\end{eqnarray}

$\chi_{xx}$ is identical to the susceptibility describing the
propagation of microwaves in an unlimited ferromagnetic medium in
Voigt geometry \cite{Camley} (propagation perpendicular to $\mathbf
M_0$). $\chi_{xx}$, $\chi_{xy}$ and $\chi_{yy}$ have the same
denominator, which becomes resonant (maximal) when $\omega =
\sqrt{\omega_r^2+\omega_r\omega_M}$. This is in accordance with the
FMR frequency of the Kittel formula for in-plane magnetized infinite
ferromagnetic films \cite{Kittel}.

This relatively simple behavior is due to the assumption, that
$\mathbf{h^{in}}$ is constant within the film stripe. This
assumption is only valid if the skin depth \cite{Guru} $\delta$ of
the microwaves in the sample is much larger than the sample
thickness. During our measurements we fix the microwave frequency
$f$ and sweep the magnetic field $H$. Consequently we find the FMR
magnetic field $H_0$ with

\begin{eqnarray}
\omega^2=\gamma^2(H_0^2+H_0M_0) \label{eqn:kittel}
\end{eqnarray}
and
\begin{equation}
H_0=\sqrt{M_0^2/4+\omega^2/\gamma^2}-M_0/2 \label{eq:h0}
\end{equation}

Now we introduce Gilbert damping\cite{Gilbert} $\alpha_G$ by setting
$\omega_r:=\omega_0-i\alpha_G\omega$ with now $\omega_0=\gamma H$
instead of $\omega_r=\gamma H$. We separate the real and imaginary
part of $\chi_{xx}$, $\chi_{xy}$ and $\chi_{yy}$:

\begin{eqnarray}
\chi_{xx}=(\omega_r\omega_M+{\omega_M}^2)\cdot F\nonumber\\
\chi_{xy}=\omega\omega_M\cdot F\\
\chi_{yy}=\omega_r\omega_M\cdot F\nonumber
\end{eqnarray}

with

\begin{eqnarray}
F=
\frac{\omega_0(\omega_0+\omega_M)-\alpha_G^2\omega^2-\omega^2+i\alpha_G\omega(2\omega_0+\omega_M)}
{(\omega_0(\omega_0+\omega_M)-\alpha_G^2\omega^2-\omega^2)^2+\alpha_G^2\omega^2(2\omega_0+\omega_M)^2}
\nonumber\\\nonumber
\approx\frac{(H+H_0+M_0)(H-H_0)+i(2H+M_0)\alpha_G\omega/\gamma}{((H+H_0+M_0)^2(H-H_0)^2+(2H+M_0)^2\alpha_G^2\omega^2/\gamma^2}
\end{eqnarray}

The approximation was done by neglecting the $\alpha_G^2\omega^2$
correction to the resonance frequency
$\omega^2=\omega_0(\omega_0+\omega_M)-\alpha_G^2\omega^2\approx\omega_0(\omega_0+\omega_M)$
what is possible if $\alpha_G\ll1$. Hence:

\begin{eqnarray}
\chi_{xx,xy,yy}\approx A_{xx,xy,yy}\cdot\frac{\Delta
H(H-H_0)+i\Delta H^2} {(H-H_0)^2+\Delta H^2} \label{eq:chixxyy}
\label{eq:chix} \label{eq:chiy}
\end{eqnarray}

with $\Delta H = ((2H+M_0)/(H+H_0+M_0))\cdot \alpha_G\omega/\gamma$.
This can be approximated as $\Delta H\approx\alpha_G\omega/\gamma$
if $|H-H_0|\ll H_0$. $A_{xx}$, $A_{xy}$ and $A_{yy}$ determine the
scalar amplitude of $\chi_{xx}$, $\chi_{xy}$ and $\chi_{yy}$.

To analyze the FMR line shape in the following, we will call the
Lorentz line shape which is proportional to $\Delta
H/((H-H_0)^2-\Delta H^2)$ symmetric Lorentz line shape and the line
shape proportional to $(H-H_0)/((H-H_0)^2-\Delta H^2)$ antisymmetric
Lorentz line shape. A linear combination of both will be called
asymmetric Lorentz line shape. $|H-H_0|\ll H_0$ allows us to
approximate:

\begin{eqnarray}
A_{xx}\approx\frac{\gamma(H_0M_0+M_0^2)}{\alpha_G\omega(2H_0+M_0)}\nonumber\\
\label{eq:AAA}
A_{xy}\approx\frac{M_0}{\alpha_G(2H_0+M_0)}\\
A_{yy}\approx\frac{\gamma H_0M_0}{\alpha_G\omega(2H_0+M_0)}
\nonumber
\end{eqnarray}

These are scalars which are independent of the DC magnetic field
$H$ and hence characteristic for the sample at fixed frequency.
Indeed the assumption of Gilbert damping is not essential for the
derivation of equation (\ref{eq:AAA}). In the event of a different
kind of damping, $\Delta H$ can also be directly input into
equation (\ref{eq:AAA}) replacing $\alpha_G\omega$. However
because of the commonness of Gilbert damping, its usage here can
provide a better feeling for the usual frequency dependence of
$A_{xx,xy,yy}$. Going ahead, equation (\ref{eq:Polder}) becomes:

\begin{eqnarray}
\mathbf m \approx \frac{\Delta H(H-H_0)+i\Delta H^2}
{(H-H_0)^2+\Delta H^2}\left(\begin{array}{ccc} A_{xx} & iA_{xy} & 0\\
-iA_{xy}& A_{yy} & 0\\
0      &    0   &0
\end{array}\right)\mathbf h
\label{eq:Atensor}
\end{eqnarray}

The $H$-field dependencies has Lorentz line shape with
antisymmetric (dispersive) real and symmetric (absorptive)
imaginary part, the amplitudes $A_{xx}$, $\pm iA_{xy}$ and
$A_{yy}$ respectively and the width $\Delta H$. Note that
$A_{xx}A_{yy}\approx A_{xy}^2$ for $|H-H_0|\ll H_0$. Consequently
the susceptibility amplitude tensor can be simplified to:

\begin{eqnarray}
\left(\begin{array}{ccc} A_{xx} & iA_{xy} & 0\\
-iA_{xy}& A_{yy} & 0\\
0      &    0   &0
\end{array}\right)\mathbf h\approx\left(\begin{array}{c} \sqrt{A_{xx}}\\
-i\sqrt{A_{yy}}\\0\end{array}\right)
\left[\left(\begin{array}{c} \sqrt{A_{xx}}\\
i\sqrt{A_{yy}}\\0\end{array}\right) \cdot\mathbf h\right]\nonumber
\end{eqnarray}
and equation (\ref{eq:Atensor}) becomes:
\begin{eqnarray}
\mathbf m=\frac{\gamma M_0}{\alpha_G\omega(2H_0+M_0)}\frac{\Delta
H(H-H_0)+i\Delta H^2} {(H-H_0)^2+\Delta
H^2}\nonumber\\\cdot\left(\begin{array}{c}
\sqrt{1+M_0/H_0}\\-i\\0\end{array}\right)\left(\left(\begin{array}{c}
\sqrt{1+M_0/H_0}\\i\\0\end{array}\right)\cdot\mathbf h\right)
\label{eq:mh}
\end{eqnarray}

It is visible that the ellipcity of $\mathbf m$ is independent of
the exciting magnetic field $\mathbf h$. Only the amplitude and
phase of $\mathbf m$ are defined by $\mathbf h$. The reason is the
weak Gilbert damping $\alpha_G$ for which much energy needs to be
stored in the magnetization precession to have a compensating
dissipation. Hence little energy input and impact from $\mathbf h$
appears.

From equation (\ref{eq:mh}) follows that $m_x$ and $m_y$ have
cardinally the ratio:

\begin{equation}
m_x/m_y=i\sqrt{1+M_0/H_0} \label{eq:mymx}
\end{equation}

Therefore $m_y$ vanishes for $\omega\rightarrow0$ and $m_x=im_y$ for
$\omega\rightarrow\infty$. This means that the precession of
$\mathbf M$ is elliptical and becoming more circular for high
frequencies and more linear (along the x-axis) for low frequencies.
This description applies for the case of an in-plane magnetized
ferromagnetic film. However in the case that the sample has circular
symmetry with respect to the magnetization direction (e.g. in a
perpendicular magnetized disc or infinite film\cite{GuiHu,Gui2}):
$\alpha_1=\beta_1$. This is the same as in the case that
$\omega\rightarrow\infty$ . Only in these cases the magnetization
precession can be described in terms of one precession cone angle
\cite{Costache06}. Otherwise distinct attention has to be paid to
$\alpha_1$ and $\beta_1$ (see \ref{magprop}). Additionally it can be
seen in equation (\ref{eq:mh}) that $\overline{m_y}/\overline{m_x}$
is also the ratio of the coupling strength of $\mathbf m$ to $h_y$
and $h_x$ respectively.

\subsection{MICROWAVE PHOTORESISTANCE}
\label{photores}

The microwave photoresistance $\Delta R_{MW}$ can be deduced from
equation (\ref{eq:RMW}).  First the microwave photovoltage is
excluded by setting the rf current $I_1=0$. Then we only regard the
microwave power dependent terms which depend on $\alpha_1$ and
$\beta_1$:

\begin{eqnarray}
\Delta R_{MW} =(U_0|_{I_1=0}-U_0|_{I_1=0,\alpha_1=0,\beta_1=0})/I_0
\nonumber\\ =R_A(-\alpha_1^2\cos2\alpha_0-\beta_1^2\cos^2\alpha_0)/2
\label{eq:Res_Alt}
\end{eqnarray}

If the magnetization lies parallel or antiparallel to the dc current
vector $\mathbf{I_0}$ along the stripe ($\alpha_0=0^\circ$ or
$\alpha_0=180^\circ$) the AMR is maximal. In this case magnetization
oscillation ($\alpha_1$ and $\beta_1$) reduces (-$\cos2\alpha_0=-1$)
the AMR by $\Delta R_{MW} =-(\alpha_1^2+\beta_1^2)R_A/2$ (negative
photoresistance). In contrast if the magnetization lies
perpendicular to $\mathbf I_0$ ($\alpha_0=90^\circ$, see figure
\ref{fig:RAlteration}) the resistance is minimal. In this case
magnetization oscillation corresponding to $\alpha_1$ will increase
($-\cos2\alpha_0=+1$) the AMR (positive photoresistance) by $\Delta
R_{MW} =+\ \alpha_1^2\cdot R_A/2$ (oscillations corresponding to
$\beta_1$ leave $\theta(t)$ constant in this case and do not change
the AMR).

The next step is to calculate $\alpha_1$ and $\beta_1$. The dc
magnetic field dependence of $\alpha_1 = |m_x|/M_0 =
|\chi_{xx}h_x+i\chi_{xy}h_y|/M_0$ and $\beta_1 = |m_y|/M_0 =
|-i\chi_{xy}h_x+\chi_{yy}h_y|/M_0$ is proportional to that of
$|\chi_{xx}|$, $|\chi_{xy}|$ and $|\chi_{yy}|$ given in equation
(\ref{eq:chixxyy}) (imaginary symmetric and real antisymmetric
Lorentz line shape). Squaring this results in symmetric Lorentz line
shape:

\begin{eqnarray}
\alpha_1^2\propto\beta_1^2\propto\left|\frac{\Delta H(H-H_0)+i\Delta
H^2} {(H-H_0)^2+\Delta H^2}\right|^2=\frac{\Delta H^2}
{(H-H_0)^2+\Delta H^2}\nonumber
\end{eqnarray}

Hence:

\begin{eqnarray}
\alpha_1^2=\frac{|A_{xx}h_x+iA_{xy}h_y|^2}{M_0^2}\cdot\frac{\Delta
H^2} {(H-H_0)^2+\Delta
H^2}\nonumber\\
\beta_1^2=\frac{|A_{yy}h_y-iA_{xy}h_x|^2}{M_0^2}\cdot\frac{\Delta
H^2} {(H-H_0)^2+\Delta H^2} \label{eq:sqr_alpha}
\end{eqnarray}

Using equation (\ref{eq:mh}) and (\ref{eq:sqr_alpha}), equation
(\ref{eq:Res_Alt}) transforms to:

\begin{eqnarray}
\label{eq:RMWcomplete} \nonumber\Delta R_{MW}=
\frac{R_A}{(\alpha_G\omega/\gamma)^2(2H_0+M_0)^2}\\
\cdot(-(H_0+M_0)\cos2\alpha_0-H_0\cos^2\alpha_0)\\\nonumber
\cdot\frac{\Delta H^2}{(H-H_0)^2+\Delta
H^2}\cdot|h_x\sqrt{H_0+M_0}+ih_y\sqrt{H_0}|^2
\end{eqnarray}

The strength of the microwave photoresistance is proportional to
$1/\alpha_G^2$. Weak damping (small $\alpha_G$) is therefore
critical for a signal strength sufficient for detection. The
magnetic field dependence shows symmetric Lorentz line shape.

The dependence of $\Delta R_{MW}$ on $\alpha_0$ in equation
(\ref{eq:RMWcomplete}) reveals a sign change and hence vanishing of
the photoresistance at
\begin{equation}
\cos^2\alpha_{0} =\frac 1 2\left(1-\frac{H_0}{3H_0+2M_0}\right)
\end{equation}
This means that the angle at which the photoresistance vanishes
shifts from $\alpha_0=\pm45^\circ$ and $\alpha_0=\pm135^\circ$ (for
$\omega\rightarrow0$) to $\alpha_0=\pm54.7^\circ$ and
$\alpha_0=\pm125.3^\circ$ respectively (for
$\omega\rightarrow\infty$) when increasing $\omega$. The reason for
this frequency dependence is the frequency dependence of the
ellipcity of $\mathbf m$ described at the end of \ref{magnet}.

\subsection{MICROWAVE PHOTOVOLTAGE}
\label{photovol}

The most obvious difference in appearence between the microwave
photoresistance discussed in paragraph \ref{photores} and the
microwave photovoltage discussed in this paragraph is that the
photoresistance is proportional to the square of the rf
magnetization (see equation (\ref{eq:Res_Alt}), $\alpha_1^2\approx
|m_x|^2/M_0^2$ and $\beta_1^2\approx |m_y|^2/M_0^2$) while the
photovoltage $U_{MW}$ is proportional to the product of the rf
magnetization and the rf current. Consequently the photovoltage has
a very different line shape: While the rf magnetization depends with
Lorentz line shape on $H$ (see equation (\ref{eq:chixxyy})), $I_1$
is independent of $H$. The line shape is hence determined by the
phase difference $\psi$ between the rf magnetization component
$\Re(m_xe^{-i\omega t})$ and the rf current $I_1\cos\omega t$. This
effect does not play a role in the case of photoresistance because
there only one phase matters namely that of the rf magnetization. In
contrast in photovoltage measurements a linear combination of
symmetric and antisymmetric Lorentz line shapes is found. This will
be discussed in detail in the following.

To isolate the microwave photovoltage in equation (\ref{eq:RMW}) the
dc current $I_0$ is set to 0:

\begin{equation}
U_{MW}=U_0|_{I_0=0}=-I_1\alpha_1\frac{ R_A\sin2\alpha_0\cos\psi}2
\label{eq:dc_voltage}
\end{equation}

From equation (\ref{eq:Polder}) we follow:

\begin{equation}
\alpha_1\cos\psi=\Re(m_x)=\Re(\chi_{xx}h_x+i\chi_{xy}h_y)
\label{eq:alphacos}
\end{equation}

We split $h_x=h_x^r+ih_x^i$ and $h_y=h_y^r+ih_y^i$ into real
($h_x^r$, $h_y^r$) and imaginary ($h_x^i$, $h_y^i$) part. This
enables us to isolate the real part in equation
(\ref{eq:dc_voltage}) using equation (\ref{eq:Atensor}):

\begin{eqnarray}
U_{MW} =\frac{I_1R_A\sin2\alpha_0}{2M_0}\cdot \label{eq:U_MW_final}
\{\frac{\left(A_{xy}h^r_y+A_{xx} h^i_x\right)\Delta
H^2}{(H-H_0)^2+\Delta H^2}\nonumber\\+\frac{\left(A_{xy}h^i_y-A_{xx}
h^r_x\right)\Delta H(H-H_0)}{(H-H_0)^2+\Delta H^2}\}
\end{eqnarray}

Conclusively in contrast to the microwave photoresistance ($\Delta
R_{MW}\propto 1/\alpha_G^2$, see equation (\ref{eq:RMWcomplete}))
the photovoltage is only proportional to $1/\alpha_G\propto
A_{xx,xy,yy}$. Thus good damping is less important for its
detection\cite{Egan}.

To understand the measurement results it will be necessary to
transform the coordinate system of equation (\ref{eq:U_MW_final}) to
(x',y,z'). In this coordinate system the rf magnetic field $\mathbf
h$ is constant during rotation as described in equation
(\ref{eq:ulorabs}).

To better understand the photovoltage line shape we have a closer
look on $\psi$: When sweeping $H$ the rf magnetization phase is
shifted by $\psi_m$ with respect to the resonance case ($H=H_0$).
The rf current has a constant phase $\psi_I$ which is defined with
respect to the magnetization's phase at resonance. The impact of the
dc magnetic field $H$ on the rf current ($I_1$, $\psi_I$) via the
FMR is believed to be negligible:

\begin{equation}
\cos\psi=\cos(\psi_m-\psi_I)=\cos\psi_m\cos\psi_I+\sin\psi_m\sin\psi_I
\label{eq:cospsi}
\end{equation}

$\psi$ is determined by the (complex) phase of $\chi_{xx}$,
$\chi_{xy}$ and $\chi_{yy}$ with respect to the resonance case
($\Re(\chi_{xy,yy})=0$ at $H=H_0$) during magnetic field sweep
(asymmetric Lorentz line shape, see equation (\ref{eq:chixxyy})):

\begin{equation}
\tan\psi_m=\frac{\Im\left(\frac{\Delta H(H-H_0)+i\Delta H^2}
{(H-H_0)^2+\Delta H^2}/i\right)}{\Re\left(\frac{\Delta
H(H-H_0)+i\Delta H^2} {(H-H_0)^2+\Delta
H^2}/i\right)}=\frac{H_0-H}{\Delta H} \label{eq:psim}
\end{equation}

It should be noted that according to the Landau-Liftshitz equation
\cite{LLG} $\mathbf h$ applies a torque on the magnetization and
hence excites $\mathbf m^t$ transversal. That is why at resonance
$m_x$ shows a phase shift of $90^\circ$ with respect to $h_x$.
Consequently in equation (\ref{eq:psim}) division by $i$ is
necessary ($\chi_{xx}$ and $\chi_{xy}$ become imaginary at
resonance).

Equation (\ref{eq:psim}) means that in case that the applied
microwave frequency is higher than the FMR frequency ($H_0>H$)
$\psi_m>0$ (note that $\mathbf m^t=\mathbf{m}e^{-i\omega t}$),
$\mathbf m^t$ is delayed with respect to the resonant case. The
other way around ($H_0<H$) the FMR frequency is higher than that of
the applied microwave field and $\mathbf m^t$ is running ahead
compared to the resonance case. Using equation (\ref{eq:psim}) we
find:

\begin{eqnarray}
\cos\psi_m=
\frac{\Delta H}{\sqrt{(H-H_0)^2+\Delta H^2}}\nonumber\\
\label{eq:cospsim}
\end{eqnarray}

Inserting equation (\ref{eq:sqr_alpha}), (\ref{eq:cospsi}),
(\ref{eq:psim}) and (\ref{eq:cospsim}) into (\ref{eq:dc_voltage})
gives:

\begin{eqnarray}
U_{MW}
=-\frac{R_AI_1\sin2\alpha_0}2\cdot\frac{|A_{xx}h_x+iA_{xy}h_y|}{M_0}
\nonumber\\\cdot\left(\frac{\Delta H^2\cos\psi_I}{(H-H_0)^2+\Delta
H^2}-\frac{(H-H_0)\Delta H\sin\psi_I}{(H-H_0)^2+\Delta H^2}\right)
\label{eq:umw}
\end{eqnarray}

The dependence on $H$ takes the form of a linear combination of
symmetric and antisymmetric Lorentz line shape with the ratio 1 :$\
\tan\psi_I$. The symmetric line shape contribution ($\propto\Delta
H$) arises from the rf current contribution that is in phase with
the rf magnetization at FMR and the antisymmetric from that
out-of-phase. This gives a nice impression of the phase $\psi_I$ of
the rf current determining the line shape of the FMR.

\subsection{VECTORIAL DESCRIPTION OF THE PHOTOVOLTAGE}
\label{vector}

To complete the discussion of the microwave photovoltage we want to
return to the approach used by Juretschke\cite{Juretschke} to
demonstrate that it is consistent with the description above. In
\ref{amr} we started with Ohm's law (scalar equation
(\ref{eq:OhmsLaw})). There we integrate an angle- and time-dependent
resistance. Here we want to start with the vectorial notation of
Ohm's law used in Juretschke's publication (equation
(1)\cite{Juretschke}). This integrates AMR and anomalous Hall effect
AHE. $\rho$ is the resistivity of the sample and $\Delta\rho$ that
additionally arising from AMR. $R_H$ is the anomalous Hall effect
constant:

\begin{equation}
\mathbf E=\rho \mathbf J+(\Delta\rho \mathbf M^2) (\mathbf J\cdot
\mathbf M)\mathbf M-R_H\mathbf J\times \mathbf M \label{eq:AMRHALL}
\end{equation}

We split $\mathbf M=\mathbf{M_0}+\mathbf m^t$ and the current
density $\mathbf J=\mathbf{J_0}+\mathbf j^t$ into their dc ($\mathbf
M_0$ and $\mathbf J_0$) and rf contributions ($\mathbf
m^t=\Re(\mathbf m e^{-i\omega t})$ and $\mathbf j^t=\mathbf
j\cos\omega t$). Constance of $|\mathbf{M}|$ allows $\mathbf m^t =
(m^t_x,m^t_y,0)$ in first order approximation only to lie
perpendicular to $\mathbf{M_0}=(0,0,M_0)$. To select the
photovoltage we set $\mathbf J_0=0$ and approximate equation
(\ref{eq:AMRHALL}) to second order in $\mathbf j^t$ and $\mathbf
m^t$.  The terms of zeroth order in both $\mathbf j^t$ and $\mathbf
m^t$ represent the sample resistance without microwave exposure and
are not discussed here. The terms of first order in either $\mathbf
j^t$ or $\mathbf m^t$ (but not both) have zero time average and do
not contribute to the microwave induced dc electric field $\mathbf
E_{MW}$. Only the terms that are simultaneously of first order in
$\mathbf j^t$ and $\mathbf m^t$ contribute to $\mathbf E_{MW}$
(compare equation (4) from Juretschke\cite{Juretschke}):

\begin{equation}
\mathbf E_{MW} =
\frac{\Delta\rho}{{M_0}^2}\left<\left(\mathbf{j}^t\mathbf{m}^t\right)\mathbf{M_0}
+\left(\mathbf{j}^t\mathbf{M_0}\right)\mathbf{m}^t\right>
-R_H\left<\mathbf{j}^t\times\mathbf{m}^t\right>
\label{eq:juretschke}
\end{equation}

The $\Delta\rho$ dependent term represents the photovoltage
contribution arising from AMR and the $R_H$ dependent term that
arising from AHE. Note that a second order of $\mathbf m^t$ appears
when applying a dc current $\mathbf{J_0}\not= 0$. It represents the
photoresistance discussed in \ref{photores}. However it we will not
be discussed here.

In the following we will calculate the photovoltage in our Permalloy
film stripe considering its geometry which fixes the current
direction. $\mathbf{j}^t=j^t_{z\prime}\mathbf{z\prime}$ along the
stripe ($\mathbf{z\prime}$ is the unit vector along the Permalloy
stripe). The small dimensions perpendicular to the stripe ($\ll L$)
will prevent the formation of a perpendicular rf current. A similar
approximation of a metal grating forming a linear polarizer has been
considered previously\cite{Gui}. The photovoltage $U_{MW}$ is also
measured along the stripe (length vector
$\mathbf{L}=\mathbf{z^\prime}\cdot2.4$ mm). When fluctuations of
$\mathbf{E}_{MW}$ along the stripe are neglected considering the
large microwave wavelength, $\lambda\approx20$ mm $\gg2.4$ mm $=L$,
we find $U_{MW}$ by multiplying $\mathbf E_{MW}$ with $\mathbf L$:

\begin{eqnarray}
\label{eqn:PVstripe}
U_{MW}=\int_0^L\mathbf{E_{MW}}\mathbf{dz^\prime} \approx\mathbf
E_{MW}\cdot\mathbf L =\nonumber\\\frac{\Delta\rho
L}{{M_0}^2}\left<j^t_{z\prime}\left(\mathbf{z\prime}\mathbf{m}^t\right)\left(
\mathbf{M_0}\mathbf{z\prime}\right)
+j^t_{z\prime}\left(\mathbf{z\prime}\mathbf{M_0}\right)
\left(\mathbf{m}^t\mathbf{z\prime}\right) \right>-0\\\nonumber
=\frac{\Delta\rho L}{M_0}\left<
j^t_{z\prime}m^t_x\right>\sin(2\alpha_0)
\end{eqnarray}

This is equivalent to equation (\ref{eq:dc_voltage}) which can be
verified by replacing $\Delta\rho j^t_{z\prime}L=R_A I_1\cos(\omega
t)$ and $m_x^t=\alpha_1M_0\cos(\omega t-\psi)$. Time averaging
results in the additional factor $\cos(\psi)/2$.

As discussed in \ref{other} the contribution belonging to the
anomalous Hall effect has no impact in this geometry because it can
only generate a photovoltage perpendicular to the rf current i.e.
perpendicular to the stripe.

Comparing our results to those of Juretschke and
Egan\cite{Juretschke,Egan}, we note that an equation similar to
(\ref{eqn:PVstripe}) has been derived in the formula for $e_{y0}$ in
equation (31) in Juretschke's publication \cite{Juretschke}. There
the photovoltage is measured parallel to the rf current as done in
our stripe. However it has to be noted that the coordinate system is
defined differently. The major difference compared to our system is
that we use a stripe shaped film to lithographically define the
direction of the rf current $I_1$, while the direction of $\mathbf
h$ is left arbitrary. In contrast to that Juretschke and
Egan\cite{Juretschke,Egan} define the direction of the rf magnetic
field and rf current by means of their microwave setup. In equation
(31) ($e_{y0}$) from Juretschke \cite{Juretschke} this results in
the additional factor $\cos\theta$ (which is equivalent to
$\cos\alpha_0$ in our work) compared to equation
(\ref{eqn:PVstripe}). This arises from the definition of $\mathbf h$
fixed parallel to the rf current (compare equation
(\ref{eq:ulorabs})).

\subsection{OTHER MAGNETORESISTIVE EFFECTS THAT COUPLE SPIN AND CHARGE CURRENT}
\label{other}

In this section we present other magnetoresistive effects which can
generate photovoltage and photoresistance like the AMR. This
selection gives a broader view on the range of effects for which the
photovoltage and photoresistance can be discussed in terms of the
analysis presented in this work. In principle every magnetoresistive
effect can modulate the sample resistance and thus rectify some of
the rf current to photovoltage.

One magnetoresistive effect is the anomalous Hall effect AHE in
ferromagnetic metals that was (together with the AMR) the basis for
the discussion of Juretschke \cite{Juretschke}. There a current with
perpendicular magnetization generates a voltage perpendicular to
both. Under microwave exposure this alternates with the microwave
frequency but in an asymmetric way due to the modulated AHE arising
from magnetization precession. The asymmetric voltage has a dc
contribution (photovoltage)\cite{Egan} which can be measured using a
2 dimensional ferromagnetic film with the magnetization neither
parallel nor perpendicular to it. The photovoltage induced by AHE
appears in the film plane perpendicular to the rf current and is
small\cite{Moller} for Permalloy (Ni$_{80}$Fe$_{20}$). Also a
photoresistive effect which alters the AHE can be expected if the
magnetization lies out-of-plane.

\begin{figure}
\centering
\includegraphics{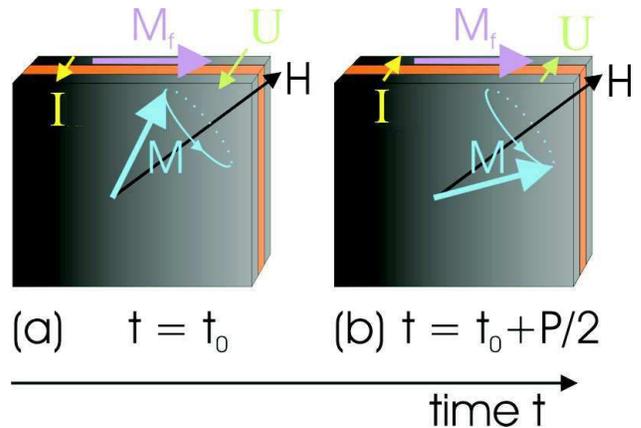}
\caption{(Color online). Microwave photovoltage in a
GMR/TMR-heterostructure (ferromagnetic($\mathbf
M$)/non-ferromagnetic/ferromagnetic($\mathbf{M_f}$)): The dynamic
magnetization $\mathbf M$ precesses (period P) in phase with the
current $\mathbf I$. (a) $\mathbf M$ lies almost perpendicular to
$\mathbf{M_f}$: high GMR/TMR. (b) $\mathbf M$ lies almost parallel
to $\mathbf{M_f}$: low GMR/TMR $\Rightarrow$ Non-zero time average
of the voltage $\mathbf U$.} \label{fig:GMRTMR}
\end{figure}

Other examples for magnetoresistive effects are GMR and TMR
structures which exhibit a photovoltage mechanism similar to that in
AMR films. The difference is that there not the direction of the
ferromagnetic layer magnetization with respect to the current
matters. Effectively instead the direction of the magnetization
$\mathbf M$ of one ferromagnetic layer with respect to that of
another layer is decisive (see figure \ref{fig:GMRTMR}). Exciting
the FMR in one layer yields again oscillation of the sample
resistance $R(t)$ and thus gives the corresponding rf voltage $U(t)$
a non-zero time average (photovoltage)
\cite{Tulapurkar,Kupferschmidt}. This is usually stronger than that
from AMR films due to the generally higher relative strength of GMR
and TMR compared to AMR.

It should be noted that in current studies of the microwave
photovoltages effect in multilayer structures, the focus is on
interfacial spin transfer effects
\cite{Tulapurkar,Sankey,Azevedo,Saitoh,Costache,Berger99,Brataas,
Wang,Kupferschmidt}.
It remains an intriguing question whether interfacial spin transfer
effects and the effect revealed in our approach based on
phenomenological magnetoresistance might be unified by a consistent
microscopic model, as Silsbee \textit{et al.} have demonstrated for
describing both bulk and interfacial spin transport under rf
excitation \cite{Silsbee}.

Multilayer structures also provide a nice example that photovoltage
generation can also be reversed when the oscillating
magnetoresistance, transforms a dc current into an rf voltage
\cite{Kiselev}, instead of transforming an rf current into a dc
voltage (photovoltage). This gives a new kind of microwave source
and seems - although weaker - also possible in AMR and AHE samples.

It can be reasoned that like microwave photovoltage the microwave
photoresistance can also be based on GMR or TMR instead of AMR: When
aligning the 2 magnetizations of both ferromagnetic layers in a GMR
or TMR structure microwave induced precession of one magnetization
is expected to increase the GMR/TMR because of the arising
misalignment with the other magnetization. With the magnetizations
initially anti parallel the opposite effect, a microwave induced
resistance decrease, is expected. Further work demonstrating these
effects would be interesting.

\section{PHOTOVOLTAGE MEASUREMENTS}
\label{voltmeas}

\subsection{MEASUREMENT SETUP}
\label{measgeom}

The sample we use to investigate the microwave photovoltage consists
of a thin (d = 49 nm) Permalloy (Ni 80\%, Fe 20\%) film stripe (200
$\mu$m wide and 2400 $\mu$m long) with 300$\times$300 $\mu$m$^2$
bond pads at both ends (see figure \ref{fig:koordinaten}). These are
connected via gold bonding wires and coaxial cables to a lock-in
amplifier. For auxiliary measurements (e.g. Hall effect) 6
additional junctions are attached along the stripe (see figure
\ref{fig:koordinaten}).

The resistance of the film stripe is $R_0+R_A=85.0\ \Omega$ for
parallel and $R_0=83.6\ \Omega$ for perpendicular magnetization.
Hence the conductance is $\sigma=1/\rho=2.9\cdot10^6\
\Omega^{-1}$m$^{-1}$ and the relative AMR is $\Delta\rho/\rho=1.7\
\%$. The absolute AMR is $R_A=1.4\ \Omega$. This is in good
agreement with previous publications \cite{Gui,GuiHu,Gui2}.

The film is deposited on a 0.5 mm thick GaAs single crystal
substrate, and patterned using photolithography and lift off
techniques. The substrate is mounted on a 1 mm polyethylene print
circuit board which is glued to a brass plate holding it in between
the poles of an electromagnet. This provides the dc magnetic field
$B = \mu_0H$ (maximal $\approx1$ T). The sample is fixed 1 mm behind
the end of a WR62 ($15.8\times7.9$ mm) hollow brass waveguide which
is mounted normal to the Permalloy film plane. The stripe is fixed
along the narrow waveguide dimension. In the K$_u$ band (12.4 - 18
GHz), that we use in our measurements, the WR62-waveguide only
transmits the TE$_{01}$ mode \cite{Guru}. The stripe was fixed with
respect to the waveguide but was left rotatable with respect to
$\mathbf H$. This allows the stripe to be parallel or perpendicular
to $\mathbf H$, but keeps the magnetic field always in the film
plane. A high precision angle readout was installed to indicate
$\alpha_0$.

\begin{figure}
\centering
\includegraphics{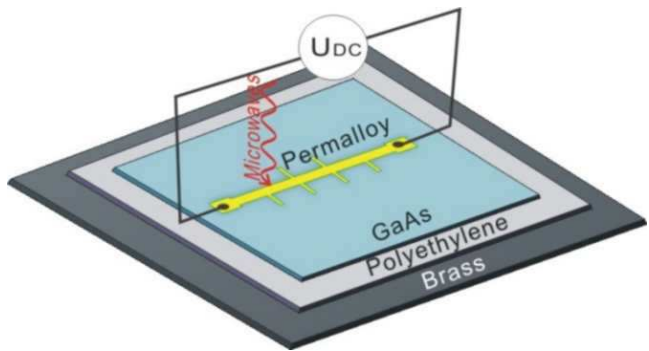}
\caption{(Color online). Sketch of the measurement geometry. A 1 mm
thick polyethylene plate is glued on a brass holder. On top of the
polyethylene a GaAs substrate is glued. On the substrate the
Permalloy (Py) stripe is defined. This is electrically wired to a
voltage amplifier for photovoltage measurements. For photoresistance
measurements an additional current source is connected parallel to
the voltage amplifier, which is not shown explicitly here. }
\label{fig:sample}
\end{figure}

The waveguide is connected to an HP83624B microwave generator by a
coaxial cable supplying frequencies of up to 20 GHz and a power of
200 mW. The power is however later significantly reduced by losses
occurring within the coaxial cable, during the transfer to the
hollow waveguide and by reflections at the end of the waveguide.
Microwave photovoltage measurements are performed sweeping the
magnetic field while fixing the microwave frequency. The sample is
kept at room temperature.

To avoid external disturbances the photovoltage was detected using a
lock-in technique: A low frequency (27.8 Hz) square wave signal is
modulated on the microwave CW-output. The lock-in amplifier,
connected to the Permalloy stripe, is triggered to the modulation
frequency to measure the resulting square wave photovoltage across
the sample. Instead of the photovoltage also the photocurrent can be
measured\cite{GuiHu}. Its strength $I_0$ can be found when setting
$U_0=0$ in equation (\ref{eq:RMW}) (instead of $I_0=0$).

\subsection{FERROMAGNETIC RESONANCE}
\label{ferro}

\begin{figure}
\centering
\includegraphics{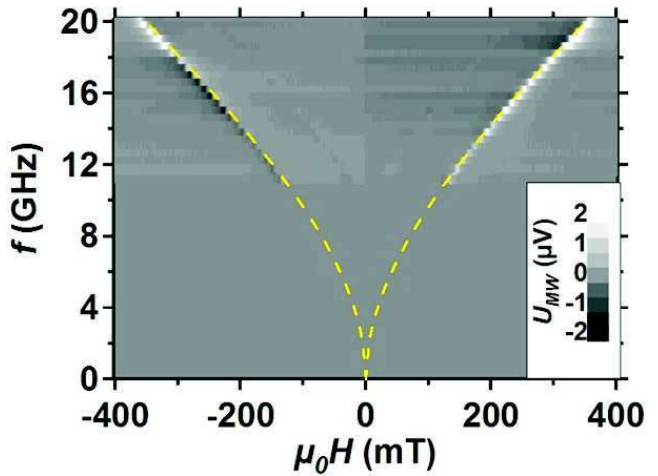}
\caption{(Color online). Gray scale plot of the measured frequency
and magnetic field dependence of the microwave photovoltage at
$\alpha_0 = 47^\circ$. The dashed line shows the calculated FMR
frequency (see equation (\ref{eqn:kittel})). The photovoltage
intensity is strongly frequency dependent because of the frequency
dependent waveguide transmission.} \label{fig:disp}
\end{figure}

The measured photovoltage almost vanishes during most of the
magnetic field sweep but shows one pronounced resonance of several
$\mu$V. The strength and line shape of this resonance are strongly
depending on $\alpha_0$ and will be discussed in \ref{lineshape}. A
line shape dependence of the photovoltage on the microwave frequency
is also found. The photovoltage with respect to the strength of the
external magnetic field $H$ and the microwave frequency
$f=\omega/2\pi$ can be seen in a gray scale plot in figure
\ref{fig:disp}, in which the resonance can be identified with the
FMR by the corresponding fits (dashed line) because the
Kittel-equation \cite{Kittel} (\ref{eqn:kittel}) for ferromagnetic
planes (our Permalloy film) applies. The magnetic parameters found
are $\mu_0M_0 \approx 1.02$ T and $\gamma \approx 2\pi\mu_0\cdot
28.8$ GHz/T. They are in good agreement with previous publications
\cite{Gui,GuiHu}.

The exact position of the FMR is obscured by its strongly varying
line shape. We overcome this problem by the productive line shape
analysis in paragraph \ref{asymm}. It is found that $H_0$ is
slightly dependent on $\alpha_0$. This can be attributed to a small
demagnetization field perpendicular to the stripes but within the
film plane arising from the finite stripe dimensions in this
direction. So, when $\mathbf{M_0}$ lies perpendicular to the stripe,
$H_0$ slightly increases compared to the value fulfilling the Kittel
equation for a plane (see equation (\ref{eqn:kittel})). In the
parallel and perpendicular case we use the approximation of our film
stripe as an ellipsoid, where we can use the corresponding Kittel
equation \cite{Kittel} (demagnetization factors $N_{x}$, $N_y$ and
$N_{z}$ with respect to the dc magnetic field):

\begin{equation}
\omega =\gamma\sqrt{\left(H_0+\left(N_{x}-N_{z}
\right)M_0\right)\left(H_0+\left(N_y-N_{z}\right)M_0\right)}
\label{eqn:kittel_gen}
\end{equation}

The difference of the resonance field between the case that $\mathbf
M_0$ lies in the film plane parallel to the stripe and perpendicular
is 1.6 mT (0.7\%) at f = 15 GHz. From this we can calculate the
small demagnetization factor $N_{x\prime} = 0.085\%$ perpendicular
to the Permalloy stripe within the film plane using equation
(\ref{eqn:kittel_gen}). From the sum rule \cite{Iwata} follows:
$N_y=1-N_{x\prime}-N_{z\prime}= 1-0.085\%-0=99.915\%$. $N_{z\prime}$
(parallel to the stripe) can be assumed to be negligibly small. This
matches roughly with the dimension of the height to width ratio (49
nm : 200 $\mu$m) of the sample. For the stripe presented in section
\ref{resmeas} similar but  stronger demagnetization effects are
found.

\label{magprop}

Now we will have a closer look on the magnetic properties of the
investigated film. Again at $f=15$ GHz we find using equation
(\ref{eq:h0}): $H_0=0.219$ T. Using asymmetric Lorentz line shape
fitting as described in \ref{asymm} we get $\alpha_G = 0.0072$.
Consequently $A_{xx}=231.1$, $A_{xy}=97.1$ and $A_{yy}=40.8$
according to equation (\ref{eq:AAA}).

Because of $\alpha_G=0.0072$ the magnetization precession does
impressive $n \approx 22$ turns before being damped to $1/e$ of its
initial amplitude ($n=1/2\pi\alpha_G$). Therefore the ellipcity of
$\mathbf m$ is almost independent of $\mathbf h$ (see paragraph
\ref{magnet}). It can be calculated from equation (\ref{eq:mymx})
that $m_x/m_y=2.38i$ at $\omega/2\pi$ = 15 GHz.

\begin{figure}
\centering
\includegraphics{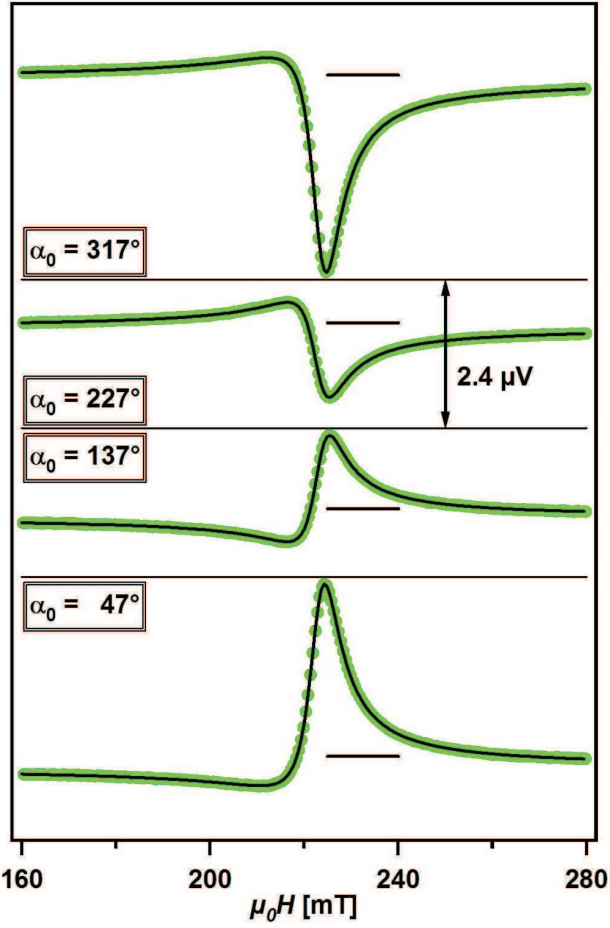}
\caption{(Color online). Fitting (black line) of the microwave
photovoltage signal (dots) for different angles $\alpha_0$ at $f=15$
GHz. The black horizontal bars indicate zero signal.}
\label{fig:measPV}
\end{figure}

\begin{figure}
\centering
\includegraphics{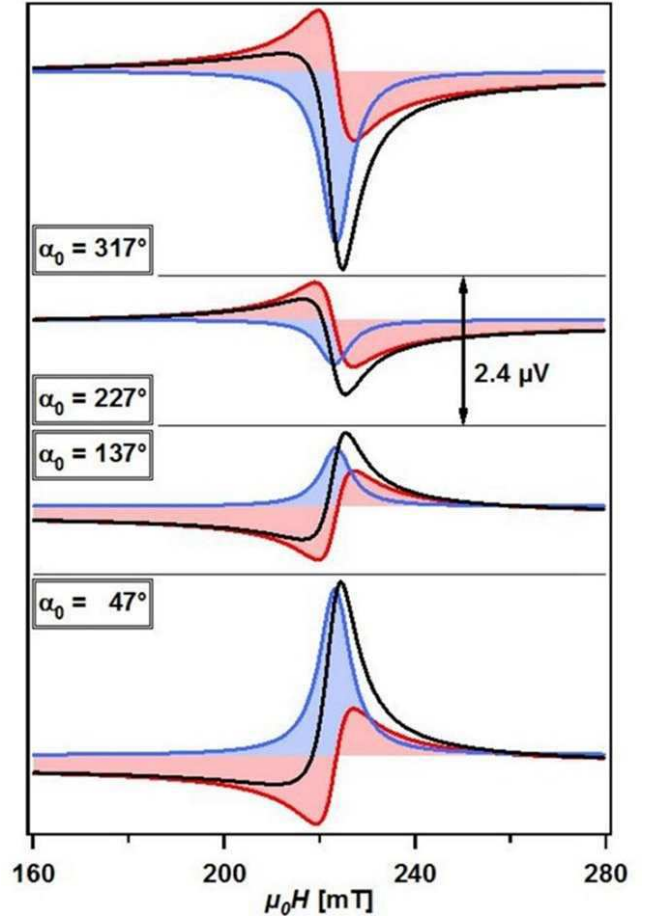}
\caption{(Color online). Symmetric and Antisymmetric contributions
to the asymmetric Lorentz line shape fit from figure
\ref{fig:measPV} (black). A small constant background is found and
added to the antisymmetric contribution.} \label{fig:pv}
\end{figure}

To check the validity of our approximation ($d\ll\delta$, see
\ref{magnet}) we will now regard the skin depth $\delta$ at $f=15$
GHz in our sample ($d=49$ nm). For $\mu=\mu_0$ (away from the FMR)
we find: $\delta=\sqrt{2/\omega\mu\rho}$ = 2.4 $\mu$m. Hence
$\delta\gg d$. This is in accordance with our approximation that
$\mathbf h$ is almost constant within the Permalloy film (see
\ref{magnet}). However in the vicinity of the FMR: $|\mu|\gg \mu_0$
and for the same frequency and conditions as above:
$\mu_L=(1+\chi_L)\mu_0=133i\mu_0$ at the FMR. Thus we approximate
$\delta_{FMR} = \sqrt{2/\omega|\mu_L|\rho}=$ 210 nm. Hence
$\delta_{FMR}$ is still significantly larger than $d$ and our
approximation is still valid.

Finally we can summerize that for samples with weak damping
($\alpha_G\ll\omega/\omega_M$) like ours the approximation $H\approx
H_0$ gives results with impressive precision (see figure
\ref{fig:measPV}) because its discrepancies are limited to the
unimportant magnetic field ranges with $|\chi_{xx}|$, $|\chi_{xy}|$,
$|\chi_{yy}|\ll1$ which are far away from the FMR.

\subsection{ASYMMETRIC LORENTZ LINE SHAPE}
\label{asymm}

Although in section \ref{ferro} the frequency dependence of the FMR
field is verified with the gray scale plot in figure 7, it is still
desirable to receive a more accurate picture of the corresponding
line shape which is found to be strongly angular dependent (see
figure \ref{fig:measPV}). In equation (\ref{eq:umw}) it is shown
that the magnetic field dependence of $U_{MW}$ exhibits asymmetric
Lorentz line shape around $H=H_0$. Hence $U_{MW}$ takes the form

\begin{eqnarray}
U_{MW}=U_{MW}^{SYM}+U_{MW}^{ANT}\nonumber\\=U_{0}^{SYM}\frac{\Delta
H^2} {(H-H_0)^2+\Delta H^2}\\\nonumber+U_{0}^{ANT}\frac{\Delta
H(H-H_0)} {(H-H_0)^2+\Delta H^2} \label{eq:Lineshape}
\end{eqnarray}

This is used to fit the magnetic field dependence of the
photovoltage in figure \ref{fig:measPV}. For clearness the symmetric
(absorptive) and antisymmetric (dispersive) contributions are shown
separately in figure \ref{fig:pv}. A small constant background is
found and added to the antisymmetric contribution. The background
could possibly arise from other weak non-resonant photovoltage
mechanisms.

The fits agree in an unambiguous manner with the measured results.
Hence they can be used to determine the Gilbert damping parameter
with high accuracy: $\alpha_G\approx\gamma\Delta
H/\omega\approx(0.72\%\pm 0.015\%$). However if the magnetization
lies parallel or perpendicular to the stripe the photovoltage
vanishes (see equation (\ref{eq:dc_voltage})). Hence we can only
verify $\alpha_G$ when the magnetization is neither close to being
parallel nor perpendicular to our stripe.

The corresponding $\alpha_G = 1/\omega\tau$ in the Nickel sample of
Egan and Juretschke\cite{Egan}, can be estimated using the
ferromagnetic relaxation time $\tau$ from their Table II. It lies in
between $\alpha_G = 0.12$ and 0.18, so being more than 16 times
higher than the value in our sample. This makes the line shape
approximation of section \ref{photovol} invalid for their case.
Consequently a much more elaborated line shape
analysis\cite{Juretschke} appears necessary.

\begin{figure}
\centering
\includegraphics{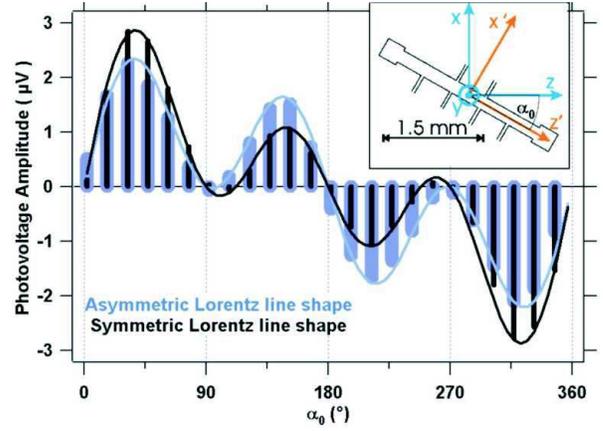}
\caption{(Color online). The Bars show the angular dependence of the
amplitude of the symmetric ($U_0^{SYM}$, thin bars) and
antisymmetric ($U_0^{ANT}$, thick bars) contribution to the
microwave photovoltage at f = 15.0 GHz. Note that both, the
symmetric and antisymmetric contribution, vanish for $\alpha_0=
0^\circ, 90^\circ, 180^\circ$ and $270^\circ$. The lines represent
the corresponding fits by means of equation (\ref{eq:FITDISP}). The
inlet shows the geometry of the investigated Permalloy stripe and
the coordinate systems from figure \ref{fig:koordinaten} (note:
$\mathbf z\parallel\mathbf H$).} \label{fig:angle}
\end{figure}

\label{lineshape}

In figure \ref{fig:measPV} the photovoltage along the stripe is
presented at 4 different angles $\alpha_0$. The signal to noise
ratio is about 1000 because of the carefully designed measurement
system, where the noise is suppressed to less than 5 nV. Because of
this good sensitivity we can verify the matching of our theory from
section \ref{theory} with the measurement results in great detail.

In the following we want to investigate the angular dependence in
detail. Therefore we transform the coordinate system of equation
(\ref{eq:U_MW_final}) according to the transformation presented in
section \ref{amr}. Doing so we can separate the contributions from
$h_{x\prime}$, $h_y$ and $h_{z\prime}$:

\begin{eqnarray}
U_{MW}\ = \frac{R_A I_1\sin(2\alpha_0)}{2M_0}\cdot\nonumber\\
\{\left(A_{xy}h_y^r+A_{xx}(h_{x\prime}^i\cos\alpha_0-h_{z\prime}^i\sin\alpha_0)
\right) \label{eq:ulorabs} \nonumber\\\cdot\frac{\Delta H^2}
{(H-H_0)^2+\Delta H^2}\\\nonumber
+\left(A_{xy}h_y^i+A_{xx}(h_{z\prime}^r\sin\alpha_0-h_{x\prime}^r\cos\alpha_0)
\right)\\\nonumber\cdot\frac{\Delta H(H-H_0)} {(H-H_0)^2+\Delta
H^2}\}
\end{eqnarray}

$h_{x\prime}$, $h_y$ and $h_{z\prime}$ are fixed with respect to the
hollow brass waveguide and its microwave configuration and do not
change when $\alpha_0$ is varied.

\begin{figure}
\centering
\includegraphics{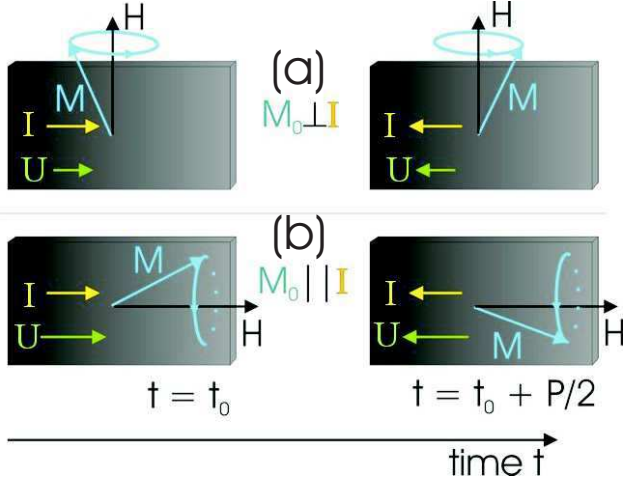}
\caption{(Color online). When the magnetic field $\mathbf H$ lies
parallel or perpendicular to the stripe, the time average voltage
vanishes. (a) $\mathbf H$ lies perpendicular to $\mathbf I$:
precession of the magnetization $\mathbf M$ leaves (after half a
period P/2) the angle $\theta$ between the axis of $\mathbf M$ and
$\mathbf I$ unchanged. Hence the AMR (and so voltage $\mathbf U$) is
also unchanged. The photovoltage vanishes. (b) $\mathbf H$ is
parallel to $\mathbf I$: $\theta$  and the AMR stay constant during
the precession of $\mathbf M$ and the time average of $\mathbf I$ is
zero. This means that only when $\mathbf H$ is neither parallel nor
perpendicular to the stripe a photovoltage is generated.}
\label{fig:dceffect2}
\end{figure}

We find that the angular dependence of the line shape in equation
(\ref{eq:ulorabs}) exhibits 2 aspects: an overall factor
$\sin(2\alpha_0)$ and individual factors ($\sin\alpha_0$,
$\cos\alpha_0$ and 1) for the terms belonging to the different
spatial components of $\mathbf h$. The overall factor
$\sin(2\alpha_0)$ arises from the AMR photovoltage mechanism and
results in vanishing of the photovoltage signal at $\alpha_0=
0^\circ, 90^\circ, 180^\circ$ and $270^\circ$. This means if
$\mathbf{M_0}$ lies either parallel, antiparallel or perpendicular
to the stripe axis. This is illustrated in figure
\ref{fig:dceffect2} and is clearly observed in our measurements (see
figure \ref{fig:angle}). We take this as a strong support for the
photovoltage being really AMR based.

Another support comes from the similarity with the planar Hall
effect \cite{Yau}. The planar Hall effect generates a voltage
$U_{PHE}$ perpendicular to the current in ferromagnetic samples
(width W) when the magnetization $\mathbf{M_0}$ lies in the
current-voltage plane. It arises as well from AMR and vanishes when
$\mathbf{M_0}$ lies either parallel or perpendicular to the current
axis.

The similarity arises because of the AMR only generating a
transversal resistance when the current is not lying along the
principle axis of its resistance matrix (parallel or perpendicular
to the magnetization). This is the same geometrical restriction as
shown above for the microwave photovoltage (see equation
(\ref{eq:dc_voltage}) and figure \ref{fig:dceffect2}).

We want to emphasize the importance that in any of these microwave
photovoltage experiments, due to the unusually strong angle
dependence, it is important to pay attention to the exact angle
adjustment of the sample with respect to the dc magnetic field
$\mathbf H$ when measuring under high symmetry conditions
($\mathbf{H}$ parallel or perpendicular to the stripe) to avoid
involuntary signal changes due to small misalignments. As found in
$90^\circ$ out-of-plane configuration \cite{GuiHu} already a
misalignment as small as a tenth of a degree can yield a tremendous
photovoltage change in the vicinity of the FMR.

\begin{figure}
\centering
\includegraphics{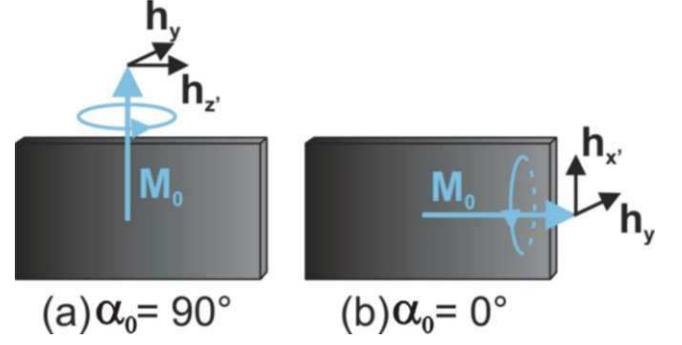}
\caption{(Color online). Angular dependent coupling of the
magnetization $\mathbf M$ to the dynamic magnetic field
$\mathbf{h}=(h_{x\prime},h_y,h_{z\prime})$. Only the components of
$\mathbf{h}$ perpendicular to $\mathbf{M_0}$ can excite precession
of $\mathbf M$ and therefore generate a dynamic $\mathbf{m}$. $h_y$
is always exciting $\mathbf{m}$. The excitation strength of
$h_{x\prime}$ and $h_{z\prime}$ is angular dependent (compare
equation (\ref{eq:ulorabs})). Here the two symmetry cases are shown:
$\mathbf M$ (a) perpendicular (only $h_{z\prime}$ and $h_y$ can
excite $\mathbf M$) and (b) parallel (only $h_{x\prime}$ and $h_y$
can excite $\mathbf M$) to the stripe. } \label{fig:mw_coupl}
\end{figure}

Finally we want to come back to the individual angular dependencies
of the photovoltage contributions arising from the different
external magnetic field components. In addition to the
$\sin(2\alpha_0)$ proportional dependence of $U_{MW}$ on $m_x$, also
the strength with which $m_x$ is excited by $\mathbf h$ depends on
$\alpha_0$. This is displayed in figure \ref{fig:mw_coupl} and
reflected by the three terms in equation (\ref{eq:ulorabs})
depending on $h_{x\prime}$, $h_{y}$ and $h_{z\prime}$ with
$\cos\alpha_0$, 1 and $\sin\alpha_0$-factors respectively. Hence the
symmetric $U_{0}^{SYM}$ and antisymmetric $U_{0}^{ANT}$ Lorentz line
shape contribution to $U_{MW}$ are fitted in figure \ref{fig:angle}
with

\begin {eqnarray}
\nonumber U_{0}^{SYM}=(U_{z\prime}^{S}\sin(\alpha_0)
+U_{x\prime}^{S}\cos(\alpha_0)+U_y^{S})\sin(2\alpha_0)\\
\label{eq:FITDISP} U_{0}^{ANT}=(U_{z\prime}^{A}\sin(\alpha_0)
+U_{x\prime}^{A}\cos(\alpha_0)+U_y^{A})\sin(2\alpha_0)
\end{eqnarray}

From $U_{z\prime}^{S}$, $U_{x\prime}^{S}$ and $U_y^{A}$ the dynamic
magnetic field components $h_{z\prime}^i$, $h_{x\prime}^i$, $h_y^i$
which are $90^\circ$ out-of-phase with respect to the rf current
$I_1$ can be determined using equation (\ref{eq:ulorabs}) and from $
U_{z\prime}^{A}$, $ U_{x\prime}^{A}$ and $U_y^{S}$ we find
$h_{z\prime}^r$, $h_{x\prime}^r$ and $h_y^r$ which are in phase with
$I_1$.

In principle $I_1$ can be separately deduced using the bolometric
effect \cite{Gui07c} as discussed in section \ref{bolo}. However for
the sample used here our usage of multiple stipes does not allow us
to address the bolometric heating to one single stripe. Consequently
the strength of $I_1$ is unknown so that we can not determine
$\mathbf h$, but only $\mathbf hI_1$.

Besides, considering the special dynamic magnetic field
configuration in our rectangular hollow waveguide no rf magnetic
field component $h_{z\prime}$ is expected to be generated along the
waveguides narrow dimension (z$^\prime$-axis) by the TE$_{01}$ mode
\cite{Guru} (which is the microwave configuration of our waveguide).
It follows that the $\sin(\alpha_0)$ terms in equation
(\ref{eq:FITDISP}) vanishes. This results in the additional symmetry
$U_{MW}(\alpha_0)=-U_{MW}(-\alpha_0)$, which is clearly observed in
our measurements (see figure \ref{fig:angle}). This symmetry was
broken when we used a round waveguide.

The vanishing of $h_{z\prime}$ in our waveguide will allow us to
plot the direction of $\mathbf h$ 2 dimensional (instead of 3
dimensional) in figure \ref{fig:rf_h}. A small deviation from the
symmetry $U_{MW}(\alpha_0)=-U_{MW}(-\alpha_0)$ is however found and
arises from a small $h_{z\prime}$ component (see table
\ref{table:table1}) which is not displayed in figure \ref{fig:rf_h}.
It might arise from the fact that the rf microwave magnetic field
$\mathbf h$ at the waveguide end already deviates from the TE$_{01}$
mode.

\subsection{DETERMINATION OF THE RF MAGNETIC FIELD DIRECTION}
\label{deter}

Using the different angular dependencies of the 3 symmetric and 3
antisymmetric terms in equation (\ref{eq:ulorabs}) $\mathbf hI_1$
can be determined. We make the assumption that the stripe itself
does not influence the rf magnetic field configuration, what is at
least the case when further reducing its dimensions. Thus the film
stripe becomes a kind of detector for the rf magnetic field $\mathbf
h$.

\begin{table}
\begin{ruledtabular}
\begin{tabular}[c]{|l|r|r|rr|r|r|}
\hline & $U^{S}_{x\prime,y,z\prime}$&$U^{A}_{x\prime,y,z\prime}$&
$A_{xx}$&$A_{xy}$&$I_1\cdot h^r_{x\prime,y,z\prime}$& $I_1\cdot
h^i_{x\prime,y,z\prime}$\\\hline
&\multicolumn{2}{c|}{($\mu$V)}&&&\multicolumn{2}{c|}{(mA$\cdot\mu$T$/\mu_0)$}\\
\hline $x^\prime$ & +2.60 &+2.55&231.1&&-15.7&+16.4\\ \hline $y$ &
+0.95&+0.30&&97.1&+14.0&+4.4 \\\hline $z^\prime$ &
+0.12&0.00&231.1&&0.0&-0.7 \\ \hline
\end{tabular}
\end{ruledtabular}
\caption{\label{table:table1} Determination of the rf magnetic field
$\mathbf h$ at the 200 $\mu$m wide stripe at 1 mm distance from the
waveguide end by means of equation (\ref{eq:ulorabs}).
$U^{S}_{x\prime,y,z\prime}$, $U^{A}_{x\prime,y,z\prime}$: Measured
amplitudes of the contributions to the symmetric and antisymmetric
Lorentz line shape of $U_{MW}$ (see equation (\ref{eq:FITDISP}))
with the angular dependence belonging to $x^{\prime}$, $y$ and
$z^\prime$ respectively (taken from the fitting in figure
\ref{fig:angle}). $A_{xx,xy}$: Corresponding amplitudes of
$\chi_{xx,xy}$. $h^r_{x\prime,y,z\prime}$,
$h^i_{x\prime,y,z\prime}$: rf magnetic field strength calculated
from $U^{S}_{x\prime,y,z\prime}$, $U^{A}_{x\prime,y,z\prime}$
(in-phase and 90$^\circ$ out-of-phase contribution with respect to
the current).}
\end{table}

\begin{figure}
\centering
\includegraphics{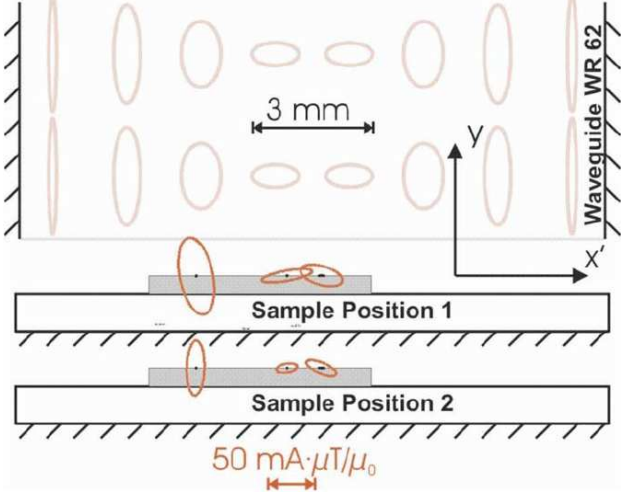}
\caption{(Color online). Direction and ellipticity of the rf
magnetic field $\mathbf h$ displayed by showing the path
$I_1\cdot\mathbf h$ passes during one cycle. This is shown at the
location of the 3 stripes (these lie normal to the picture on top of
the gray GaAs-substrate; the 200 $\mu$m wide stripe to the right)
for two sample positions. $I_1\cdot\mathbf h$ was determined by
means of equation (\ref{eq:ulorabs}). The upper right path
corresponds to the $I_1\cdot\mathbf h$ from table
\ref{table:table1}. The hatched edges indicate metal surfaces
reflecting microwaves. Within the waveguide the rf magnetic field
$\mathbf h$ corresponding to the TE$_{10}$-mode is displayed in the
background.} \label{fig:rf_h}
\end{figure}

To test this an array of 36 additional 50 $\mu$m wide and 20 $\mu$m
distant Permalloy stripes of the same height and length as the 200
$\mu$m wide stripe described above (see section \ref{measgeom}) was
patterned beside this one. The 50 $\mu$m wide stripes were connected
with each other at alternating ends to form a long meandering
stripe\cite{Gui}. Four stripes were elongated on both ends to
300$\times$300 $\mu$m$^2$ Permalloy contact pads. For the outer two
stripes and the single 200 $\mu$m stripe $\mathbf hI_1$ is
calculated from the measured photovoltage using equation
(\ref{eq:U_MW_final}). Table \ref{table:table1} shows the measured
voltage and the corresponding $\mathbf h I_1$ for the 200 $\mu$m
stripe at 1 mm distance from the waveguide. $\mathbf hI_1$ for all 3
stripes is displayed in figure \ref{fig:rf_h}, while positioning the
sample at 2 distances (1 and 3.5 mm respectively) from the waveguide
end. For comparison the rf magnetic field $\mathbf h$ configuration
of the TE$_{01}$-mode is displayed in the background. From other
measurements we can estimate that $I_1$ lies somewhere in the 1
mA-range.

It is worth noting that possible inhomogeneities of the rf magnetic
field $\mathbf h$ within the Permalloy stripes will be averaged
because $U_{MW}$ is linear in $\mathbf h$. Determining the sign of
the rf magnetic field components from the photovoltage contributions
signs exhibits a certain complexity because a lot of attention has
to be paid to the chosen time evolution ($e^{i\omega t}$ or
$e^{-i\omega t}$) and coordinate system (right hand or left hand).
However the sign only reflects the phase difference with respect to
the rf current. The rf current is admittedly not identical for
different stripe positions. Consequently the comparison of the rf
magnetization phase at different stripe locations is obscured.

It is a specially interesting point concerning microwave
photovoltage that the phase of the individual components of the rf
magnetic field with respect to the rf current, and therefore also
with respect to each other can be determined. The phase information
is encoded in the line shape, which is a particular feature of the
microwave photovoltage described in this work.

At this point only determining $\mathbf hI_1$ is possible because
$I_1$ is unknown. However in paragraph \ref{bolo} an approach to
determine $I_1$ using the bolometric effect is presented. Using this
approach the bolometric photoresistance is the perfect supplement
for the photovoltage. It delivers unknown $I_1$ with almost no
additional setup.

\section{PHOTORESISTANCE MEASUREMENTS}
\label{resmeas}

The principle difficulties when detecting the AMR induced
photoresistance are to increase the microwave power for a sufficient
signal strength and to reduce the photovoltage signal, which is in
general much stronger and superimposes with the photoresistance. We
overcome the microwave power problem by using high initial microwave
power (316 mW) and a coplanar waveguide (CPW) \cite{GuiHu} which
emits the microwaves as close as possible to the Permalloy film
stripe ($0.137\times20\times2450\ \mu$m$^3$) with which we detect
the photoresistance. Its resistance is found to be $R=880\ \Omega$
and the AMR $R_A=15\ \Omega$. Its magnetic properties ($\gamma$,
$M_0$) are almost identical to that of the sample investigated in
section \ref{voltmeas}. We use again lock-in technique like in
\ref{measgeom} with now an additional dc current from a battery to
measure resistance instead of voltage. The strong microwave power
results in strong rf currents within the sample which give a
specially strong photovoltage signal (see equation (\ref{eq:umw})).
To achieve a sufficiently strong photoresitance signal the dc
current $I_0$ and rf current $I_1$ have to be increased to the
maximal value that does not harm the sample (a few mA, hence
$I_0\approx I_1$).

Ignoring the trigonometric factors $\sin2\alpha_0$, $\cos2\alpha_0$
and $\cos\psi$ as well as the photoresistance term depending on
$\beta_1$ (that is always smaller than $\alpha_1$) the photovoltage
signal ($U_{MW}=\alpha_1\sin(2\alpha_0)\cos\psi R_AI_1/2$, equation
(\ref{eq:dc_voltage})) and the photoresistance signal ($\Delta
R_{MW}I_0\approx-\alpha_1^2\cos(2\alpha_0)R_AI_0/2$,
(\ref{eq:Res_Alt})) become almost identical. But the major
difference is that the photoresistance is multiplied by $\alpha_1^2$
and the photovoltage only by $\alpha_1$. As $\alpha_1$ is
particularly small ($<1^\circ$) in our experiments, this means that
$\Delta R_{MW}I_0$ is much smaller than $U_{MW}$. However
suppressing $U_{MW}$ is possible because it vanishes for
$\alpha_0=0^\circ,\ 90^\circ,\ 180^\circ,\ 270^\circ$ (see equation
(\ref{eq:dc_voltage})). A very precise tuning of $\alpha_0$ with an
accuracy below $0.1^\circ$ is necessary to suppress $U_{MW}$ below
$\Delta R_{MW}I_0$. Fortunately in contrast to $|U_{MW}|$, $|\Delta
R_{MW}|$ is maximal for $\alpha_0=0^\circ,\ 90^\circ,\ 180^\circ,\
270^\circ$. In the following we will first discuss the bolometric
photoresistance arising from microwave heating of the sample and
afterwards the AMR induced photoresistance that is discussed above.

\subsection{BOLOMETRIC (NON-RESONANT)}
\label{bolo}

\begin{figure}
\centering
\includegraphics{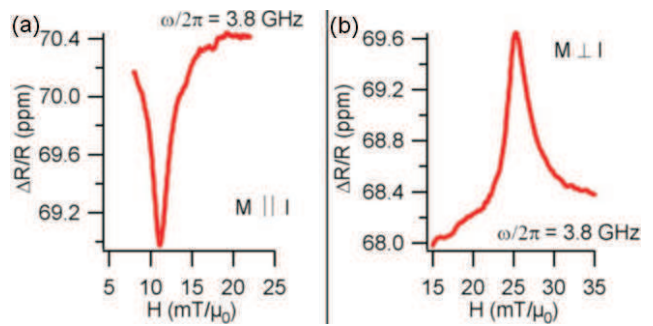}
\caption{(Color online). Photoresistance $\Delta R_{MW}$ measurement
(stripe resistance $R$). The curves show the difference between the
signals $\Delta  U$ with $I_0=+5$ mA and $I_0=-5$ mA at $P=316$ mW:
$\Delta R=(\Delta U(I_0=+5$ mA$)-\Delta U(I_0=-5$mA$))/10$ mA. The
subtraction suppresses the photovoltage dependence on absolute
$|I_0|$ (for example from bolometric AMR change). For both curves
the dc magnetic field $\mathbf H$ (and so $\mathbf M$) was applied
within the film plane, but for a) parallel to the stripe (and hence
to the dc current $I_0$) and for b) perpendicular. A non-resonant
background of about 70 ppm from bolometric photoresistance is found.
It is decreases by about 1.2 ppm when the sample is turned from
parallel to perpendicular configuration. This is caused by the 1.7
\% AMR which changes $R$ and the bolometric signal proportionally.
The FMR signal has almost Lorentz line shape and its position is
significantly changing when the sample is turned from parallel to
perpendicular position (see section \ref{res_pr}). }
\label{fig:prmeas}
\end{figure}

The AMR-induced $\Delta R_{MW}$ is not the only photoresistive
effect present in our Permalloy film stripe. Also non-resonant
heating by the microwave rf current $I_1$ results in a (bolometric)
photoresistance. The major difference compared to the AMR-based
photoresistance is that the bolometric photoresistance is almost
independent of the applied dc magnetic field $H$ and that its
reaction time to microwave exposure is much longer (in the order of
ms) than that of the AMR-based photresistance ($1/\alpha_G\omega$,
in the order of ns) \cite{Gui07c}. The non-resonant bolometric
photoresistance is found with a typical strength of $(\Delta
R/R)/P=$ 0.2 ppm/mW (see figure \ref{fig:prmeas}).

The bolometric heating power $P_{bol}$ arises from resistive
dissipation of the rf current $I_1$ in the sample
($P_{bol}=\left<RI^2\right>=RI_0^2+RI_1^2/2$). This can hence be
used to determine $I_1$, which is otherwise an unknown in equation
(\ref{eq:umw}). $I_1$ can be determined for example by finding the
corresponding dc current $I_0$ with the same bolometric resistance
change. However, especially in the sample we use the thermal
conductivity of the GaAs-crystal on which our Permalloy stripes were
deposited is so high (55 W$/$m$\cdot$K) that the different stripes
are strongly thermally coupled. Thus we can not address the
bolometric signal of one stripe solely to the rf current of the same
stripe. This effect was verified comparing the resistance changes
from one stripe while applying a dc current through an other stripe.
Hence determination of $|I_1|$ by means of equation (\ref{eq:umw})
is only possible when using a substrate material with low heat
conductance (e.g. glass) or by not depositing more than one stripe.

\begin{figure}
\centering
\includegraphics{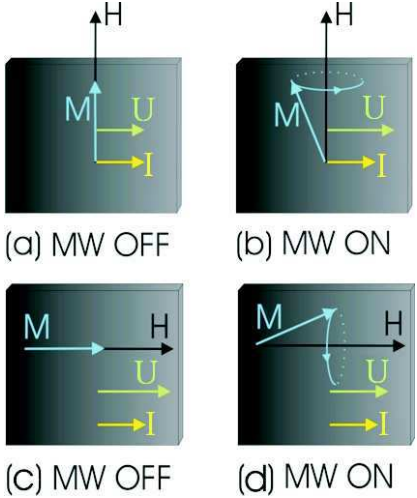}
\caption{(Color online). Demonstration of the angular dependence of
the microwave photovoltage: Without microwaves (a,c) the AMR is
minimal in perpendicular configuration (a) of $\mathbf M$ and
$\mathbf I$ and maximal in parallel configuration (c). When the
microwaves are switched on the resistance increases in parallel
configuration (b) and decreases in perpendicular configuration (d).
} \label{fig:prsketch}
\end{figure}

\subsection{AMR BASED (RESONANT)}
\label{res_pr}

In contrast to the non-resonant bolometric photoresistance in
\ref{bolo}, the typically 50 times weaker resonant AMR-based
photoresistance is very hard to detect. After visualizing it by
using the CPW and turning the sample into a high symmetry position
(parallel or perpendicular to $\mathbf H$) it is still necessary to
regard the difference of the photoresistance measured with the same
current strength but with reversed current sign instead of measuring
with only one current direction. This eliminates the remaining still
significant photovoltage signal, which depends on the absolute
current strength possibly due to bolometric AMR change.

Measurement results are presented in figure \ref{fig:prmeas} for
$f=3.8$ GHz. There it can be seen that (as deduced in
\ref{photores}), if the stripe lies parallel to the magnetization,
the AMR is maximal and the resistance decreases when the FMR is
excited (negative photoresistance). In contrast in the perpendicular
case the AMR is minimal and we measure a resistance increase
(positive photoresistance). This behavior is schematically explained
in figure \ref{fig:prsketch}. The curves in figure \ref{fig:prmeas}
show the photoresistance at the FMR with symmetric Lorentz line
shape as predicted in \ref{photores}.

Using equation (\ref{eqn:kittel}) we calculate $\mu_0H_0=16.6$ mT.
However a deviation of $H_0$ is found in both, parallel
($\mu_0H_0=11.1$ mT) and perpendicular ($\mu_0H_0=25.3$ mT),
configuration. This is due to demagnetization which gives rise to an
FMR shift with respect to the result from the infinite film
approximation (compare equation (\ref{eqn:kittel_gen})). $N_x=0.7\%$
can be assumed because of this shift.

Using equation (\ref{eq:mymx}) we find that for our conditions
$m_x/m_y=7.9i$. Consequently we can neglect the contribution from
$\beta_1=|m_y|/M_0$ in equation (\ref{eq:Res_Alt}) and find
$|m_x|=13$ mT using $\Delta R_{MW}=(\Delta R/R)\cdot R=1.23$
m$\Omega$ (from figure \ref{fig:prmeas}) and thus
$\alpha_1=\sqrt{2\cdot\Delta R_{MW}/R_A}=0.73^\circ$ and
$\beta_1=\alpha_1/|m_x/m_y|=0.09^\circ$. The smallness of
$\beta_1$ is the reason for the resonant photoresistance strength
being almost identical for $\mathbf M
\parallel \mathbf I$ and $\mathbf M \perp \mathbf I$ (although the
sign is reversed). We must expect $|m_x|$, $\alpha_1$ and $\beta_1$
to be even a little bit larger due to our lock-in measurement
technique only detecting the sinusoidal contribution to the square
wave signal from the microwaves.

The photoresistive decrease is in accordance with that found by
Costache \emph{et al.} \cite{Costache06}. There the magnetization is
aligned with the current ($\alpha_0=0$). Thus applying an rf
magnetic field decreases the AMR from $R_A$ to $R_A\cos^2\theta_c$.
This is used to determine the precession cone angle $\theta_c$ by
assuming $\theta_c = \alpha_1 = \beta_1$.

The height to width ratio of the strip is 35 nm to 300 nm. Because
of the magnetization lying along the stripe,\cite{Costache06} the
magnetization precession strongly deviates from being circular.
Using the corresponding parameters $\mu_0M_0=1.06$ T,
$\gamma=2\pi\mu_0\cdot28$ GHz/T and $\omega/2\pi=10.5$ GHz), we find
from equation (\ref{eq:mymx}) that the ratio of the amplitudes is
$m_x/m_y= 3.15i$. This indicates strongly elliptical precession and
suggests that distinguishing $\alpha_1$ and $\beta_1$ would provide
a refined description compared to that using the cone angle
$\theta_c$, as discussed in paragraph \ref{photores}.

\section{CONCLUSIONS}
\label{conclu}

We have presented a comprehensive study of dc electric effects
induced by ferromagnetic resonance in Py microstrips. A theoretical
model based on a phenomenological approach to magnetoresistance is
developed and compared with experiments. These provide a consistent
description of both photovoltage and photoresistance effects.

We demonstrate that the microwave photoresistance is proportional to
the square of magnetization precession amplitude. In the special
case of circular magnetization precession, the photoresistance
measures its cone angle. In the general case of arbitrary sample
geometry and elliptical precession, we refine the cone angle concept
by defining 2 different angles, which provide a precise description
of the microwave photoresistance (and photovoltage) induced by
elliptical magnetization precession. We show that the microwave
photoresistance can be either positive or negative, depending on the
direction of the dc magnetic field.

In contrast to the microwave photoresistance, we find that the
microwave photovoltage is proportional to the product of the
in-plane magnetization precession component with the rf current.
Consequently it is sensitive to the magnetic field dependent phase
difference between the rf current and the rf magnetization. This
results in a characteristic asymmetric photovoltage line shape,
which crosses zero when the rf current and the in-plane component of
the rf magnetization are exactly 90$^\circ$ out of phase. Therefore,
the microwave photovoltage provides a powerful insight into the
phase of magnetization precession, which is usually difficult to
obtain.

We demonstrate that the asymmetric photovoltage line shape is
strongly dependent on the dc magnetic field direction, which can be
explained by the directional dependence of the magnetization
precession excitation. By using the model developed in this work,
and by combining such a sensitive geometrical dependence of the
microwave photovoltage with the bolometric photoresistance which
independently measures the rf current, we are now in a position to
detect and determine the external rf magnetic field vector, which is
of long standing interest with significant potential applications.

\section*{ACKNOWLEDGEMENTS}

We thank G. Roy, X. Zhou and G. Mollard for technical assistance and
D. Heitmann, U. Merkt, and the DFG for the loan of equipment. N. M.
is supported by the DAAD. This work has been funded by NSERC and
URGP grants awarded to C.-M. H.


\begin{thebibliography}{199}

\bibitem{Guru} B. S. Guru and H. R. Hiziroglu,
\emph{Electromagnetic Field Theory Fundamentals, Second Edition}
(Cambridge University Press, 2004).

\bibitem{Gurney} B. A. Gurney \emph{et al.}, J. Shi, and R. R. Katti,
in \emph{Ultrathin Magnetic Structures IV} edited by B. Heinrich and
J.A.C. Bland (Springer, Berlin, 2004), Chapter 6,7 and 8.

\bibitem{Zhu} J.-G. Zhu, and Y. Zheng, in \emph{Spin Dynamics in Confined
Magnetic Structures I} edited by B. Hillebrands and K. Ounadjela
(Spinger, Berlin, 2002), P. 289-323.

\bibitem{Tulapurkar} A. A. Tulapurkar \emph{et al.}, Nature
$\mathbf{438}$, 339 (2005)
\bibitem{Sankey} J. C. Sankey \emph{et al.}, Phys. Rev. Lett.
$\mathbf{96}$, 227601 (2006).

\bibitem{Azevedo}
A. Azevedo,  L. H. Vilela Leão, R. L. Rodriguez-Suarez, A. B.
Oliveira, and S. M. Rezende, J. Appl. Phys. {\bf 97}, 10C715 (2005).

\bibitem{Saitoh}
E. Saitoh, M. Ueda, H. Miyajima, and G. Tatara, Appl. Phys. Lett.
{\bf 88}, 182509 (2006).

\bibitem{Costache}
M.V. Costache, M. Sladkov, S. M. Watts, C. H. van der Wal, and B. J.
van Wees, Phys. Rev. Lett. {\bf 97}, 216603 (2006); J. Grollier, M.
V. Costache, C. H. van der Wal, and B. J. van Wees, J. Appl. Phys.
{\bf 100}, 024316 (2006).

\bibitem{Gui} Y. S. Gui, S. Holland, N. Mecking, and C.-M.
Hu, Phys. Rev. Lett. $\mathbf{95}$, 056807 (2005).

\bibitem{GuiHu} Y. S. Gui, N. Mecking, X. Zhou, Gwyn Williams, and C.-M. Hu,
Phys. Rev. Lett. $\mathbf{98}$ 107602 (2007).

\bibitem{Gui2} Y. S. Gui, N. Mecking, and C.-M. Hu, Phys.
Rev. Lett. $\mathbf{98}$, 217603 (2007).

\bibitem{Gui07c} Y. S. Gui, N. Mecking, A. Wirthmann,
L. H. Bai, and C.-M. Hu, Appl. Phys. Lett. $\mathbf{91}$, 082503
(2007).

\bibitem{Costache06} M.V. Costache, S. M. Watts, M. Sladkov, C. H. van der Wal, and B.
J. van Wees, Appl. Phys. Lett. {\bf 89}, 232115 (2006)

\bibitem{Yamaguchi} A. Yamaguchi, H. Miyajima, T.Ono,
Y. Suzuki, S. Yuasa, A. Tulapurkar, and Y. Nakatani, Appl. Phys.
Lett. $\mathbf{90}$, 182507 (2007); \emph{ibid}, $\mathbf{90}$,
212505 (2007).

\bibitem{Oh} Dong Keun Oh, \textit{et al.} J. Magn. Magn. Mater. {\bf 293}, 880
(2005); Je-Hyoung Lee and Kungwon Rhie, IEEE Trans. on Mag. {\bf
35}, 3784 (1999).

\bibitem{Goennenwein} S. T. Goennenwein, S. W. Schink, A. Brandlmaier, A. Boger, M.
Opel, R. Gross, R. S. Keizer, T. M. Klapwijk, A. Gupta, H. Huebl, C.
Bihler, and M. S. Brandt, Appl. Phys. Lett. $\mathbf{90}$, 162507
(2007).

\bibitem{Slonczewski}
J.C. Slonczewski, J. Magn. Magn. Mater. {\bf 159}, L1 (1996).

\bibitem{Berger96}
L. Berger, Phys. Rev. B {\bf 54}, 9353 (1996).

\bibitem{Berger99}
L. Berger, Phys. Rev. B {\bf 59}, 11465 (1999).

\bibitem{Brataas}
A. Brataas, Yaroslav Tserkovnyak, Gerrit E. W. Bauer, and Bertrand
I. Halperin, Phys. Rev. B {\bf 66}, 060404(R) (2002).

\bibitem{Wang}
Xuhui Wang, Gerrit E. W. Bauer, Bart J. van Wees, Arne Brataas, and
Yaroslav Tserkovnyak, Phys. Rev. Lett. {\bf 97}, 216602 (2006).

\bibitem{Hu}
C.-M. Hu, C. Zehnder, Ch. Heyn, and D. Heitmann, Phys. Rev. B {\bf
67}, 201302(R) (2003).

\bibitem{Juretschke} H. J. Juretschke,
J. Appl. Phys. $\mathbf{31}$, 1401 (1960).

\bibitem{Silsbee}
R.H. Silsbee, A. Janossy, and P. Monod, Phys. Rev. B {\bf 19}, 4382
(1979).

\bibitem{Moller} W. M. Moller, and H. J. Juretschke, Phys. Rev. B
$\mathbf{2}$, 2651 (1970).

\bibitem{Polder} D. Polder, Philosophical Magazine
$\mathbf{40}$, 99 (1949).

\bibitem{Gilbert} T. L. Gilbert, IEEE Trans. on Magn. $\mathbf{40}$, 3443
(2004).

\bibitem{LLG} L. Landau, and L. Liftshitz, Physik Z. Sowjet.
$\mathbf{8}$ 153 (1935).

\bibitem{Camley} R. E. Camley, and D. L. Mills, J. Appl.
Phys. $\mathbf{82}$, 3058 (1997).

\bibitem{Kittel} C. Kittel, Phys. Rev. $\mathbf{73}$, 155
(1948).

\bibitem{Egan} W. G. Egan, and H. J. Juretschke, J. Appl. Phys.
$\mathbf{34}$, 1477 (1963).

\bibitem{Kupferschmidt} J. N. Kupferschmidt, Shaffique Adam,
and P. W. Brouwer, Phys. Rev. B $\mathbf{74}$, 134416 (2006).

\bibitem{Kiselev} S. I. Kiselev, J. C. Sankey, I. N. Krivorotov, N. C. Emley,
R. J. Schoelkopf, R. A. Buhrman, and D. C. Ralph, Nature
$\mathbf{425}$, 380 (2003).

\bibitem{Iwata} M. S. Sodha and N. C. Srivasta \emph{Microwave
Propagation in Ferrimagnets} (Plenum Press, New York, 1981).

\bibitem{Yau} K. L. Yau, and J. T. H. Chang, J. Phys. F
$\mathbf 1$, 38 (1971).

\end{thebibliography}
\end{document}